\documentclass[rmp,twocolumn,superscriptaddress,showpacs,floatfix]{revtex4}

\usepackage{graphicx}

\newcommand{\be}{\begin{equation}}
\newcommand{\ee}{  \end{equation}}
\newcommand{\ba}{\begin{eqnarray}}
\newcommand{\ea}{  \end{eqnarray}}
\newcommand{\ve}{\varepsilon}

\begin{document}

\title{Random Matrices and Chaos in Nuclear Spectra}

\author{T.~Papenbrock}
\affiliation{Department of Physics and Astronomy, University of Tennessee,
Knoxville, TN~37996, USA}
\affiliation{Physics Division,
Oak Ridge National Laboratory, Oak Ridge, TN 37831, USA}
\author{H.~A.~Weidenm\"uller}
\affiliation{Max--Planck--Institut f\"ur Kernphysik,
D-69029 Heidelberg, Germany}

\begin{abstract}
We speak of chaos in quantum systems if the statistical properties of
the eigenvalue spectrum coincide with predictions of random--matrix
theory. Chaos is a typical feature of atomic nuclei and other
self--bound Fermi systems. How can the existence of chaos be
reconciled with the known dynamical features of spherical nuclei?
Such nuclei are described by the shell model (a mean--field theory)
plus a residual interaction. We approach the question by using a
statistical approach (the two--body random ensemble): The matrix
elements of the residual interaction are taken to be random variables.
We show that chaos is a generic feature of the ensemble and display
some of its properties, emphasizing those which differ from standard
random--matrix theory. In particular, we display the existence of
correlations among spectra carrying different quantum numbers. These
are subject to experimental verification.
\end{abstract}

\maketitle

\section{Random Matrices and Chaos}

\subsection{Introduction}

This paper contains a colloquium talk on the application of random
matrices to nuclear spectroscopy. We discuss the origin of
random--matrix theory (RMT), some of its predictions, and its relation
to classical chaos. We show that the predictions of RMT often agree
very well with spectroscopic data in nuclei. In the main part of the
paper, we address the question: How can this success of RMT be
reconciled with our knowledge of the dynamical behavior of nuclei as
embodied in the nuclear shell model?  Although actually formulated for
nuclei, we believe that our arguments apply likewise in modified form
to other Fermi systems like atoms and molecules.

\begin{figure}[h]
\vspace{5 mm}
\centerline{\includegraphics[width=0.34\textwidth]{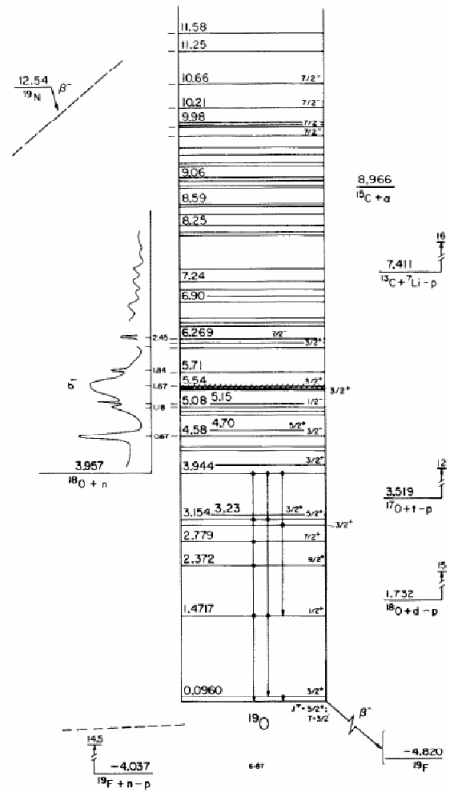}}
\vspace{3 mm}
\caption {The spectrum of $^{19}$0. Taken from Ref.~\cite{Ajz87}.}
\label{fig1}
\end{figure}

Near the ground state, the spectra of self--bound Fermi systems are
essentially discrete. As an example, Fig.~\ref{fig1} shows the
measured spectrum of the nucleus $^{19}$O. Up to the threshold for
decay into a neutron (n) and $^{18}$O (the ``neutron threshold'') at
an excitation energy of 3.957 MeV, the levels shown decay only by
gamma emission; the ground state is unstable against beta decay. The
long lifetimes of these decay modes render the levels virtually
discrete. Above neutron threshold, the levels acquire finite particle
decay widths (which typically are much larger than those for beta and
gamma decay) and are also seen as resonances in cross sections (like
in the one for the reaction n + $^{18}$O). The levels in
Fig.~\ref{fig1} carry quantum numbers which reflect symmetries of the
nuclear Hamiltonian: Total spin $J$ corresponds to rotational
invariance, total isospin $T$ to proton--neutron symmetry, parity $P$
to mirror reflection symmetry. In heavy nuclei, the Coulomb
interaction between protons breaks the isospin quantum number. In its
general appearance, Fig.~\ref{fig1} is representative not only for
nuclei but also for atoms, molecules, and atomic clusters.

\subsection{Regular Motion}

\begin{figure}[h]
\vspace{5 mm}
\centerline{\includegraphics[width=0.4\textwidth]{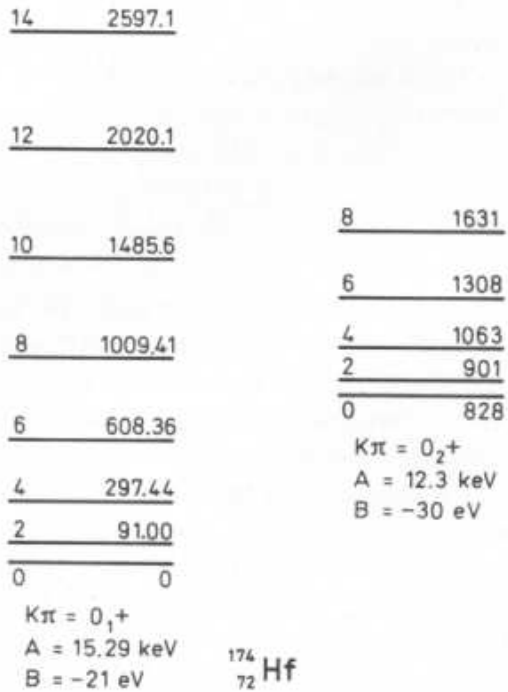}}
\vspace{3 mm}
\caption {Rotational bands in $^{174}$Hf. Taken from Ref.~\cite{Boh}.}
\label{fig2}
\end{figure}

The excitation energies and widths for beta, gamma, and particle decay
of the levels found experimentally can often be successfully described
in terms of simple integrable dynamical models. In analogy to
classical mechanics, we then speak of regular motion. A striking
example is furnished by the existence of rotational bands. These are
characterized by a spin/parity $J^P$ sequence $0^+, 2^+, 4^+, \ldots$
and excitation energies proportional to $J(J+1)$. They correspond to a
rotation of the entire (non--spherical) nucleus about some
axis. Fig.~\ref{fig2} shows two such bands in the nucleus
$^{174}$Hf. Another example for regular motion is provided by the
independent--particle model: nucleons move independently in the mean
field of the nucleus. This model corresponds to the most elementary
version of the nuclear shell model and accounts for the existence of
``magic numbers'' for neutrons and protons (i.e., numbers where a
major shell is filled and where the nuclear binding energy attains
maxima) as well as for spins and parities of ground states of nuclei
differing by one unit in proton or neutron number from closed--shell
nuclei. We discuss the nuclear shell model in more detail in the next
Section.

\subsection{Non--regular Motion}

The emphasis in this paper is on chaotic motion in nuclei. The
previous Subsection serves as a reminder that there is strong evidence
for regular motion in nuclei, especially in the vicinity of the ground
state. We now turn to the equally strong evidence for chaotic
motion. We put the argument in the form of a semi--historical
narrative which will occupy the remainder of this Subsection.

\begin{figure}[h]
\vspace{5 mm}
\centerline{\includegraphics[width=0.4\textwidth]{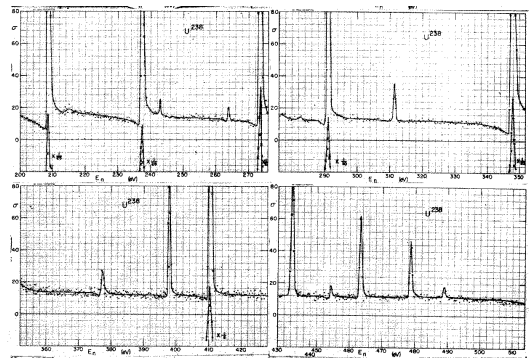}}
\vspace{3 mm}
\caption {Total cross section for scattering of neutrons by $^{238}$U
versus neutron energy. Taken from Ref.~\cite{Rain}.}
\label{fig3}
\end{figure}

The first evidence for chaotic motion in nuclei came from
spectroscopic data on levels at neutron threshold, i.e., rather far
from the ground state. In general, it is exceedingly difficult to
unambiguously identify the positions, spins, and parities of such
levels. Neutron time--of--flight spectroscopy offers the unique
opportunity to acquire precise spectroscopic information in a range of
excitation energies which is otherwise virtually inaccessible. The
neutron threshold in medium--mass and heavy nuclei typically occurs at
an excitation energy of 8 to 10 MeV (and not at $\approx$ 4 MeV as in
$^{19}$O, see Fig.~\ref{fig1}). Measuring the total cross section for
slow neutrons versus kinetic energy, one observes resonances. Each of
the resonances corresponds to a nuclear state at an excitation energy
of around 8 or 10 MeV. The ground states of even--even nuclei have
spin/parity $0^+$. Slow neutrons carry angular momentum zero and have
spin $1/2$. For even--even target nuclei, the states seen, therefore,
all have spin/parity $1/2^+$. We show here not the earliest but some
of the best data. In the 1970s, the Columbia group around Rainwater
measured time--of--flight spectra of slow neutrons scattered on a
number of heavy nuclei. For each target nucleus, they observed a
sequence of up to 200 $1/2^+$ levels. (That number was limited by the
resolution of the time--of--flight spectrometer.) Fig.~\ref{fig3}
shows their data for the total neutron scattering cross section on
$^{238}$U versus neutron energy.  We see a number of very narrow
resonances (widths typically smaller than 1 eV) with typical spacings
of 10 eV. The energies of the associated levels were determined with
the help of an $R$--matrix multi--level analysis~\cite{Lane}.

\begin{figure}[h]
\vspace{5 mm}
\centerline{\includegraphics[width=0.4\textwidth]{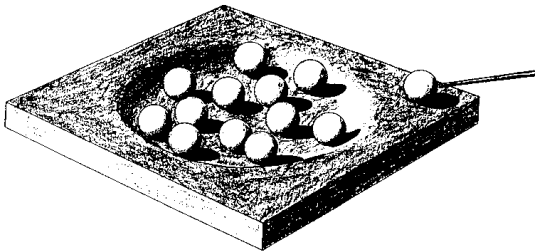}}
\vspace{3 mm}
\caption {Wooden toy model simulating N. Bohr's compound nucleus.
Taken from Ref.~\cite{NBohr1}.}
\label{fig4}
\end{figure}

Already in the 1930s, similar data (of much inferior quality) had led
N. Bohr to formulate his ``compound nucleus hypothesis''~\cite{NBohr}.
Bohr argued that the existence of narrowly spaced narrow resonances is
incompatible with independent--particle motion in the nucleus. Indeed,
simple estimates using a central potential yield single--particle
$s$--wave level spacings around 1 MeV (and not 10 eV as shown in
Fig.~\ref{fig3}), and $s$--wave decay widths for neutron--instable
states of around 100 keV (and not $<$ 1 eV as shown in
Fig.~\ref{fig3}). Bohr argued that the existence of such narrow
resonances could not be understood without assuming strong
interactions between the incident neutron and the nucleons in the
target. A wooden toy model (shown in Fig.~\ref{fig4}) was to
demonstrate this idea of the compound nucleus. The billiard balls
represent nucleons, the queue indicates the kinetic energy of the
incident neutron, and the trough simulates the attractive mean
field. Since the publication of N. Bohr's 1937 paper and until the
discovery of the nuclear shell model in 1949, the idea of
independent--particle motion in the nucleus was virtually inacceptable
in the nuclear physics community. And as we shall see, understanding
chaos in nuclei is almost synonymous to reconciling N.  Bohr's idea of
the compound nucleus with the nuclear shell model.

\subsection{Random Matrices}

When Wigner~\cite{Wign} introduced random matrices into physics, he
did not refer explicitly to N. Bohr's 1937 paper~\cite{NBohr}.
Nevertheless we believe that his work was motivated by and can be seen
as the mathematical formulation of the idea of the compound nucleus.
It is noteworthy that the work started in 1951, two years after the
discovery of the shell model.

How can one deal with a situation where the constituents of a quantum
system interact strongly? We suppose that we deal with a
non--integrable system without remaining symmetries. By that we mean
that extant symmetries of the Hamiltonian like total spin and parity
have been taken into account; the Hamiltonian has block--diagonal
form; the individual blocks are free of further symmetries; we focus
attention on one such block to which we refer for simplicity as the
Hamiltonian. Is it possible to make generic statements about the
spectrum and eigenvalues of such a system? Following
Wigner~\cite{Wign}, we consider the matrix representation $H_{\mu
\nu}$ of the Hamiltonian in Hilbert space. Here $\mu, \nu = 1, \ldots,
N$ and $N \gg 1$ (we let $N \to \infty$ at the end of the
calculation). Nuclei are time--reversal invariant. Therefore, we can
choose a representation where $H_{\mu \nu}$ is real and symmetric.
There are no further symmetries beyond rotational symmetry and parity.
Now comes the decisive and unusual step: Rather than considering the
individual Hamiltonian of the actual physical system, we study an
entire ensemble of Hamiltonians all having the same symmetry. It turns
out that while we cannot make generic statements about the individual
Hamiltonian, such statements are possible about almost all members of
the ensemble.

The ensemble is defined in such a way that it combines generality with
the symmetry of $H$: It consists of real and symmetric matrices. To
avoid a preferred direction in Hilbert space, it is chosen to be
invariant under those transformations in Hilbert space which preserve
the symmetry of $H$.  These are the orthogonal transformations. The
ensemble is defined in terms of a probability density in matrix
space. That density has the form
\be
C \exp \bigg( - \frac{N \ {\rm trace} ( H^2 )}{\lambda^2} \bigg)
\prod_{\mu \leq \nu} {\rm d} H_{\mu \nu}
\label{1}
\ee
The last factor is the product of the differentials of the independent
matrix elements. That factor is orthogonally invariant, and so is the
trace of $H^2$ in the exponent. The Gaussian factor is introduced
because a cutoff is needed to render the integrals over matrix space
convergent.  The ensemble is characterized by a single parameter
$\lambda$, which has the dimension of energy. This parameter
determines the mean level spacing of the ensemble, i.e., the mean
value of the distance between two neighboring eigenvalues. We will be
interested in the {\it fluctuations} of the spacings of neighboring
levels around that mean value, and in the correlations between
spacings of different pairs of neighboring levels.  These are
characterized by certain local fluctuation measures which are
introduced below. All measures depend on the level spacing expressed
in units of the mean level spacing and are, therefore, independent of
the parameter $\lambda$ and altogether parameter--free. The factor $C$
is a normalization factor. The factor $N$ in the exponent (with $N$
the matrix dimension) guarantees that the spectrum has finite range.
Expression~(\ref{1}) defines the Gaussian orthogonal ensemble (GOE) of
random matrices. Dyson~\cite{dys} showed that there are three
canonical random--matrix ensembles defined by symmetry: the GOE, the
Gaussian unitary ensemble (GUE) (for time--reversal non--invariant
systems), and the Gaussian symplectic ensemble (GSE) (for systems with
half--integer spin which are not rotationally invariant). GUE, GOE,
and GSE carry the Dyson parameters $\beta = 1,2,$ and $4$,
respectively.

While the symmetry of the Hamiltonian and, thus, the form of the
invariant measure in expression~(\ref{1}) is dictated by quantum
mechanics, the choice of the Gaussian cutoff is arbitrary and dictated
by convenience. It has been shown, however, that an entire class of
different choices (like, for instance, taking the trace of the fourth
power of $H$ in the exponent) yields the same results for the local
fluctuation measures. This important property is referred to as
universality of the ensemble. It guarantees that the local fluctuation
measures predicted by the GOE are generic.

Random--matrix theory (RMT) only predicts ensemble averages of
observables (calculated by integrating the observable over the
ensemble~(\ref{1})). But experimentally, we always deal with a single
Hamiltonian, not an ensemble.  The applicability of RMT to a single
system is guaranteed by the property of ergodicity: For almost all
members of the ensemble, the ensemble average of an observable (as
given by RMT) is equal to the running average of that observable taken
over the spectrum of a single member. Ergodicity guarantees that the
quantitative and parameter--free predictions of RMT can be
meaningfully compared with experimental data.

\subsection{Fluctuation Measures}

\begin{figure}[h]
\vspace{5 mm}
\centerline{\includegraphics[width=0.4\textwidth,angle=270]{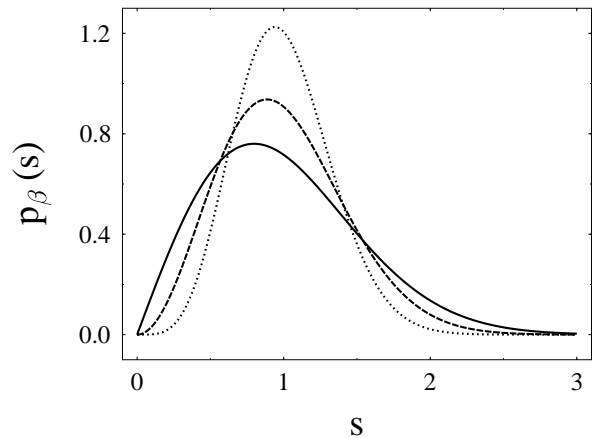}}
\vspace{3 mm}
\caption {The nearest--neighbor spacing distribution $P_\beta(s)$
versus $s$, the actual spacing in units of the mean level spacing,
for the three canonical ensembles with $\beta = 1,2,4$ (solid, dashed,
and dotted lines). $\beta = 2$ corresponds to the GOE. Taken from
Ref.~\cite{TG}.}
\label{fig5}
\end{figure}

The two fluctuation measures most frequently employed in analyzing
data are the nearest--neighbor spacing (NNS) distribution and the
Dyson--Mehta or $\Delta_3$ statistic. Fig.~\ref{fig5} gives the NNS
distribution $P_\beta(s)$, i.e., the probability distribution of
spacings, versus $s$ (the actual spacing of two neighboring
eigenvalues in units of their mean spacing) for Dyson's three
canonical ensembles. The distributions are characterized by level
repulsion at short distances ($P_\beta(s) \propto s^\beta$ for small
$s$) and a Gaussian falloff for large $s$. We note that if the system
were integrable, all eigenvalues would carry different quantum
numbers, and level repulsion were absent. The $\Delta_3$ statistic
measures correlations between eigenvalue spacings. Let ${\cal N}(E)$
denote the total number of levels below energy $E$. Clearly ${\cal
N}(E)$ is a step function which increases by one unit as $E$ passes an
eigenvalue.  The $\Delta_3$ statistic measures how well ${\cal N}(E)$
is on average approximated by a straight line. It is defined by
\be
\Delta_3(L) = \bigg\langle {\rm min}_{a, b} \frac{1}{L}
\int_{E_0}^{E_0 + L} \ \bigg( {\cal N}(E') - a E' - b \bigg)^2 {\rm d}
E' \bigg\rangle_{E_0} \ .
\label{2}
\ee
Minimization with respect to $a$ and $b$ determines the best straight
line. The angular brackets denote an average over the initial energy
$E_0$ and over the ensemble. All energies are in units of the mean
level spacing, and $\Delta_3(L)$ is, therefore, parameter free. If the
spacings were totally uncorrelated, then $\Delta_3$ would be linear in
$L$. It would grow much less fast with $L$ if the spacings were
correlated in such a way that a large spacing is always followed by a
small one and vice versa. The actual behavior of $\Delta_3(L)$ is
shown as a solid line in Fig.~\ref{fig8}: $\Delta_3(L)$ grows
essentially logarithmically with $L$. This property is referred to as
the ``stiffness'' of RMT spectra.

For our later discussion, it is important to note that a GOE spectrum
does not carry any information content beyond the mean level
spacing. This is obvious since any realization of $H$ is obtained by
drawing random numbers from the Gaussian distribution~(\ref{1}). Thus,
it would be useless to apply spectroscopic analyses to such a
spectrum. Put differently, a GOE Hamiltonian has $N(N+1)/2$
independent matrix elements. Counting shows that we need to measure
all $N$ eigenvalues and all $N$ eigenfunctions to reconstruct
$H$. This is very different for typical dynamical systems where the
Hamiltonian is known except for a small number of parameters.

\subsection{Quantum Chaos}

\begin{figure}[h]
\vspace{5 mm}
\centerline{\includegraphics[width=0.4\textwidth]{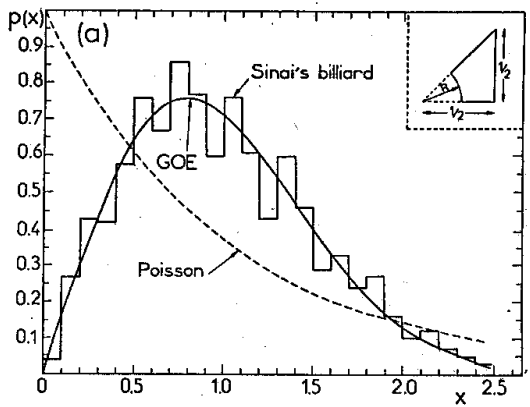}}
\vspace{3 mm}
\caption {The NNS distribution for the Sinai billiard (histogram) and
the GOE prediction (solid line). Taken from Ref.~\cite{Bohi}.}
\label{fig7}
\end{figure}

\begin{figure}[h]
\vspace{5 mm}
\centerline{\includegraphics[width=0.4\textwidth,angle=270]{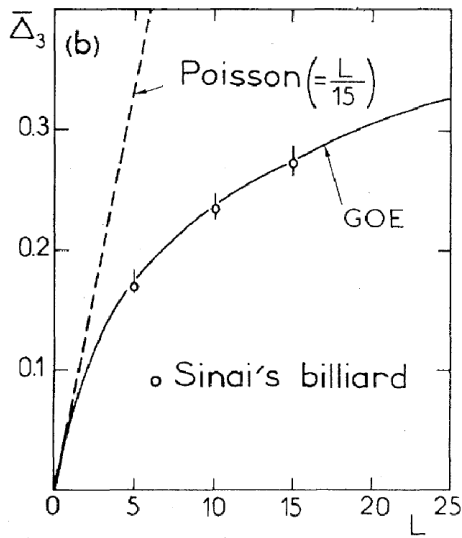}}
\vspace{3 mm}
\caption {The $\Delta_3$ for the Sinai billiard (histogram) and the
GOE prediction (solid line). Taken from Ref.~\cite{Bohi}.}
\label{fig8}
\end{figure}

Suppose the fluctuation measures of an experimental spectrum agree
with GOE predictions. What does such a result imply physically? The
understanding of this question was much advanced by the study of
few--degrees--of--freedom quantum systems which are chaotic in the
classical limit. The development culminated in the
Bohigas--Giannoni--Schmit conjecture~\cite{Bohi}. It says: {\it The
spectral fluctuation properties of a quantum system which is chaotic
in the classical limit, coincide with those of the canonical
random--matrix ensemble that has the same symmetry}.  The authors of
Ref.~\cite{Bohi} supported their conjecture by calculating numerically
the NNS distribution and the $\Delta_3$ statistic for the Sinai
billiard, a system which is time--reversal invariant and fully chaotic
in the classical limit, and comparing the result with the GOE
prediction. This is shown in Figs.~\ref{fig7} and \ref{fig8}.

Agreement like the one shown in Figs.~\ref{fig7} and \ref{fig8} has
since been found in many other chaotic systems. Moreover, an
analytical proof of the Bohigas--Giannoni--Schmit conjecture has
recently been published~\cite{heusler}. That proof uses the
semiclassical approximation and generic properties of chaotic
trajectories. We are, thus, led to consider agreement between the
spectral fluctuations of a physical system and RMT predictions as a
signal for chaotic motion. It is in that sense that we will use the
word chaos also in nuclei. This is done with a caveat: The
semiclassical proof in Ref.~\cite{heusler} is not applicable directly
to nuclei because nuclei are too dense to admit a semiclassical
approximation.

Quantum chaos has been studied intensely during the last 20 years or
so~\cite{haake}. These studies have focused mainly on dynamical
systems with few degrees of freedom. The Sinai billiard is a prime
example. Here chaos is due to the fact that the inner circle and the
outer rectangle possess incommensurate symmetries. In
contradistinction, the atomic nucleus is a many--body system. We will
show that in such a system, the dynamical features which cause chaos
are distinctly different from the ones which cause chaos in
few--degrees--of--freedom systems.

\subsection{Chaos in Nuclei}
\label{cin}

\begin{figure}[h]
\vspace{5 mm}
\centerline{\includegraphics[width=0.4\textwidth]{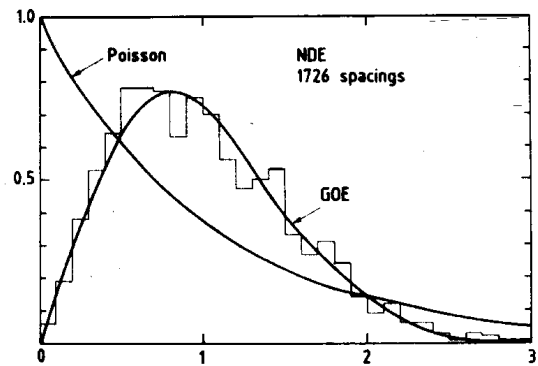}}
\vspace{3 mm}
\caption {The NNS distribution for the nuclear data ensemble
(histogram) and the GOE prediction (solid line). Taken from
Ref.~\cite{Haq}.}
\label{fig9}
\end{figure}

\begin{figure}[h]
\vspace{5 mm}
\centerline{\includegraphics[width=0.4\textwidth,angle=270]{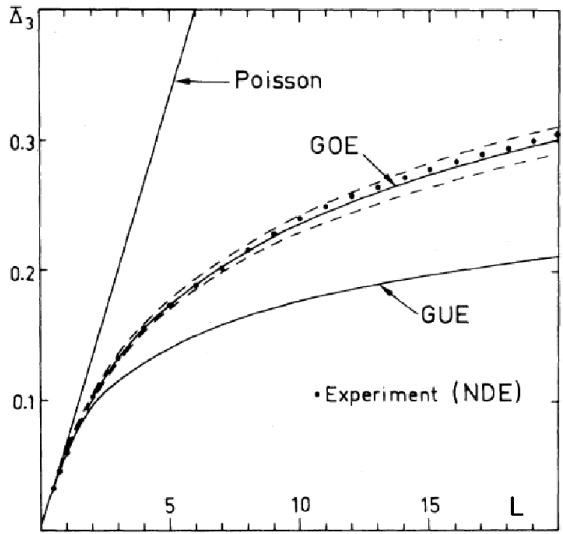}}
\vspace{3 mm}
\caption {The $\Delta_3$ for the nuclear data ensemble (data points)
and the GOE prediction (solid line). Taken from Ref.~\cite{Haq}.}
\label{fig10}
\end{figure}

Now we are in the position to return to the data shown in
Fig.~\ref{fig3}.  Combining the information on the excited levels in
$^{239}$U obtained from these data with that of many similar
experiments on the scattering of slow neutrons and of protons at the
Coulomb barrier by other nuclei, Haq, Pandey, and Bohigas~\cite{Haq}
formed the ``nuclear data ensemble''.  It consists of 1726
spacings. These were used to determine the NNS distribution and the
$\Delta_3$ statistic. The results are shown in Figs.~\ref{fig9} and
\ref{fig10}. The solid lines labeled ``Poisson'' correspond to
totally uncorrelated levels and, thus, to integrable systems. The
agreement with the GOE prediction is impressive. We conclude that at
neutron threshold (and at the Coulomb barrier for protons) nuclei
display chaotic motion. The deep physical insight of N. Bohr, who
conceived the idea of the compound nucleus and who designed the wooden
model of Fig.~\ref{fig4}, is borne out by the data: Idealized models of
billiards are prime examples of chaotic motion, and nuclear levels at
neutron threshold display chaotic behavior. Since 1983, there has been
growing evidence for chaotic motion in nuclei also at lower excitation
energies, although the statistics of such data is usually not as good
as near neutron threshold.

In conclusion, there is strong evidence for chaotic motion in nuclei.
How can this evidence be reconciled with the extant information on
nuclear dynamics (the shell model)? And -- given the fact that GOE
spectra carry zero information content beyond the mean level spacing --
which is the information content of nuclear spectra in the regime of
chaotic motion? These are the questions we address in the remainder of
this paper.

\section{Dynamical Aspects}

\subsection{Chaos in the Nuclear Shell Model}

\begin{figure}[h]
\vspace{5 mm}
\centerline{\includegraphics[width=0.4\textwidth]{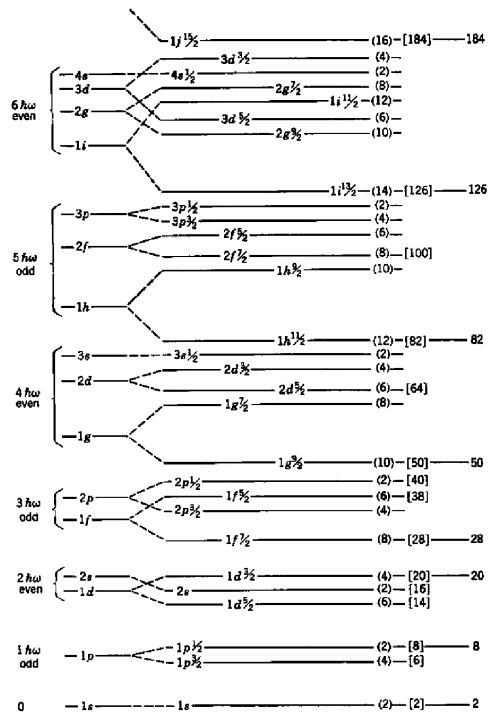}}
\vspace{3 mm}
\caption {Level sequence in the nuclear shell model. Taken from
Ref.~\cite{May}.}
\label{fig11}
\end{figure}

The shell model is the universal model for the structure of atoms and
nuclei. In nuclei, it consists of a central potential $U$ with a
strong spin--orbit interaction (the ``mean field'') and a fairly weak
``residual interaction'' $V$ which accounts for the remaining part of
the nucleon--nucleon interaction. The central potential $U$ defines
the major shells shown in Fig.~\ref{fig11}: The $1s$--shell, the
$1p$--shell, the $2s1d$--shell, etc. in spectroscopic notation, the
latter with the subshells $s_{1/2}$, $d_{3/2}$ and $d_{5/2}$ where the
indices denote the nucleon spin. The subshells have different
single--particle energies and the states within a major shell are,
therefore, not totally degenerate. Still, large degeneracies remain
when the shell contains more than a single nucleon. These degeneracies
are lifted by the residual interaction $V$. That interaction conserves
parity and is rotationally invariant and invariant under time
reversal. It is weak in the sense that it produces configuration
mixing primarily within the same major shell. More precisely: The
magnitude of a typical matrix element is small compared with the
spacing of the centroids of adjacent major shells but comparable with
that of adjacent subshells in the same major shell. In all that
follows we will, therefore, consider only a single major shell. And
the residual interaction is predominantly a two--body interaction, $V
= \sum_{\alpha < \beta} V(\vec{r}_\alpha, \vec{r}_\beta)$ although
there is evidence that three--body forces are needed to obtain
quantitative agreement with experimental spectra. Again, we will
confine ourselves for simplicity to two--body forces.

The Hamiltonian of the shell model is
\be
H = \sum_\alpha \left(-{\hbar^2\over 2m}\Delta_\alpha +
U(\vec{r}_\alpha)\right) +\sum_{\alpha < \beta} V(\vec{r}_\alpha,
\vec{r}_\beta) \ .
\ee
Here, $\vec{r}_\alpha$ denotes the coordinates and spin/isospin
degrees of freedom of the nucleon labeled $\alpha$. The
single--particle part of $H$ has eigenstates $| j \rangle$ and
eigenvalues $\ve_j$. We use second quantization and write $H$ as
\be
\label{second}
H = \sum_j \varepsilon_j a^\dagger_j a_j + {1\over 4}\sum_{ijkl}
v(ij;kl) a^\dagger_i a^\dagger_j a_l a_k \ . 
\ee
Here, $a_j^\dagger$ creates a nucleon in a single-particle state
$j$. The two--body matrix elements $v(ij;kl)$ represent the residual
interaction. In writing Eq.~(\ref{second}) we have not paid attention
to conserved quantum numbers like spin and isospin. This was done in
order to keep the notation simple. More details concerning this aspect
are given in Section~\ref{tbre}.

In the shell model it is assumed that all shells but one are
completely filled, and attention is focused on that last shell (the
``valence shell''). Distributing the valence nucleons over the various
subshells in the valence shell, one constructs a basis of
antisymmetrized many--body states $| J \mu \rangle$. We use a
short--hand notation where $J$ stands for total spin, total isospin,
and parity, while $\mu$ is a running index. These are used to
calculate the matrix elements
\be
H_{\mu \nu}(J) = \langle J \mu | H | J \nu \rangle
\label{3}
\ee
of the Hamiltonian~(\ref{second}) in the subspace with quantum numbers
$J$. Diagonalization of $H_{\mu \nu}(J)$ yields the eigenvalues and
eigenfunctions. The latter are used to calculate various transition
matrix elements. This scheme has been used with considerable success
in many parts of the periodic table. Exceptions are mainly deformed
nuclei which occur in the middle between large major shells. When we
refer to Eq.~(\ref{second}) in the text further below, we always do so
with the understanding that that equation refers to the valence shell
only.

In the sequel we focus attention to the $2s1d$--shell (in short: the
$sd$--shell). The $1s$--shell and the $1p$--shell are filled, taking a
total of 16 nucleons. Thus, the $sd$--shell describes nuclei with mass
numbers between 16 (O) and 40 (Ca). We do so for practical reasons:
The dimensions of the Hamiltonian matrices in the $sd$--shell are
quite manageable (maximal dimensions are of the order of $10^3$) while
much larger numbers occur in higher shells. We sometimes consider also
a single $j$--shell with half--integer $j$. Although not realistic in
the framework of the nuclear shell model, this idealization is useful
for purposes of orientation.

\begin{figure}[h]
\vspace{5 mm}
\centerline{\includegraphics[width=0.4\textwidth]{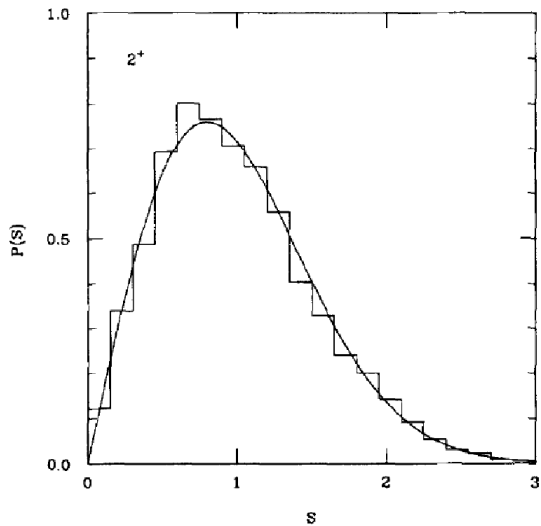}}
\vspace{3 mm}
\caption {NNS spacing distribution for the $J = 2, T = 0$ states
of 12 nucleons in the $sd$--shell (histogram) compared to the GOE
(line). Taken from Ref.~\cite{Zel}.}
\label{fig12}
\end{figure}

\begin{figure}[h]
\vspace{5 mm}
\centerline{\includegraphics[width=0.4\textwidth]{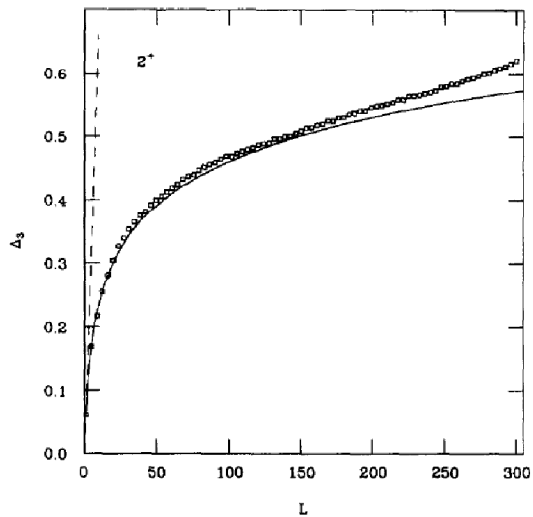}}
\vspace{3 mm}
\caption {$\Delta_3$ statistic for the $J = 2, T = 0$ states
of 12 nucleons in the $sd$--shell (data points) compared to the 
GOE (solid line). Taken from Ref.~\cite{Zel}.}
\label{fig13}
\end{figure}

Which are the fluctuation properties of the resulting spectra? That
question was studied, for instance, in Ref.~\cite{Zel}. The authors
used standard single--particle energies and a standard two--body
residual interaction to calculate the positions of the states with
spin $J = 2$ and isospin $T = 0$ for 12 nucleons in the
$sd$--shell. The matrix dimension is 874. In Figs.~\ref{fig12} and
\ref{fig13}, the NNS spacing distribution and the $\Delta_3$ statistic
for these levels are compared with GOE predictions. The agreement is
very good. (It would be desirable to confirm this result
experimentally. Unfortunately, the data needed are not available:
There is no more than a handful of levels known for each value of spin
and isospin; such levels cluster in the vicinity of the ground
state). In Ref.~\cite{Zel} numerous further results (on thermodynamic
properties, on the statistics of eigenfunctions etc.) were
described. While it was found that the input used in shell--model
calculations (the distribution of many--body matrix elements) differs
from that used in the GOE, the results agree to a very large extent
with GOE predictions.

The result displayed in Figs.~\ref{fig12} and \ref{fig13} and the other
results obtained on chaos in shell--model calculations are surprising
and call for a deeper analysis. Indeed, the Hamiltonian of the shell
model is quite different from a GOE matrix. And yet, it is able to
produce GOE--like spectral fluctuations! It does not seem unreasonable
to speculate that the shell model would also be able to account for the
GOE--like spectral fluctuations shown in Figs.~\ref{fig9} and
\ref{fig10} (this cannot be checked because of the huge dimensions of
the matrices involved). At the same time, the shell model accounts well
for some of the regular features observed experimentally like magic
numbers and spectroscopic data on low--lying states. The questions
raised at the end of Subsection~\ref{cin} must, therefore, be
reformulated as follows: Are the results shown in Figs.~\ref{fig12} and
\ref{fig13} a generic property of the shell model or are they due to
very special circumstances? In other words: How is chaos produced in
the shell model? And what is the information content of shell--model
spectra? A further question would be: How does the shell model manage
to produce both regularity and chaos? We will not address that last
question. Suffice it to say that regularity is mostly seen when we
consider states with different quantum numbers like in Fig.~\ref{fig2},
while chaos is manifest in long sequences of states carrying identical
quantum numbers. The way we use them here, regularity and chaos are not
neccessarily contradictory concepts.

\subsection{Two--body Random Ensemble (TBRE)}
\label{tbre}

To answer these questions, it is neccessary to go somewhat more deeply
into the details of shell--model calculations. The single--particle
states in the shell model carry the half--integer spin quantum number
$j$. The two--body interaction scatters a pair of nucleons in states
$(j_1, j_2)$ into states $(j_3, j_4)$ which may or may not be
identical to $(j_1, j_2)$. The two--body interaction conserves total
spin, total isospin, and parity. Coupling the initial and the final
pair of states to total spin $s$ (and suppressing the analogous
coupling for isospin for simplicity), we write the antisymmetrized
reduced matrix elements of the two--body interaction in the form
$\langle j_3 j_4 s | V | j_1 j_2 s \rangle$ or, more simply, as
$v(\alpha)$ where the index $\alpha$ labels different two-body matrix
elements. (The word ``different'' refers to matrix elements not
connected by symmetry properties.) Counting (which must include spin
and isospin) shows that in the $sd$--shell, $\alpha$ ranges from $1$
to $63$, and in a single $j$--shell with half--integer $j$, from $1$
to $j + 1/2$.

Chaos has to do with the complete mixing of states in Hilbert space.
How does the residual interaction accomplish such mixing? We focus
attention on the two--body interaction and assume that all subshells
belonging to a major shell are degenerate. (It is intuitively obvious
and has been shown numerically that the mixing of shell--model states
becomes weaker as the differences between single--particle energies of
subshells become comparable to the magnitudes of the $v(\alpha)$.) We
accordingly omit the single--particle energies in Eq.~(\ref{second})
altogether. The Hamiltonian $H_{\mu \nu}$ of Eq.~(\ref{3}) is then
completely determined by the two--body interaction in
Eq.~(\ref{second}). Thus, it has the form
\be
H_{\mu \nu}(J) = \sum_\alpha v(\alpha) \ C_{\mu \nu}(J, \alpha) \ .
\label{4}
\ee
The sum over $\alpha$ is equivalent to the sum over $\{ijkl\}$ in
Eq.~(\ref{second}). Each term in the sum in Eq.~(\ref{4}) is the
product of two contributions. The $v(\alpha)$ stand for the matrix
elements denoted by $v(ij;kl)$ in Eq.~(\ref{second}) (except that we
have now paid proper attention to the coupling of angular momenta) and
represent the specific features of the two--body interaction. Except
for angular momentum coupling coefficients, the elements of the
coefficient matrices $C_{\mu \nu}(J, \alpha)$ are the matrix elements
of the operators $a^\dagger_i a^\dagger_j a_l a_k$ in
Eq.~(\ref{second}). These matrices are determined by the major shell
we are in, by the coupling scheme we have chosen to construct the
states $| J \mu \rangle$, and by the specific two--body interaction
operator labeled $\alpha$ which defines the matrix element
$v(\alpha)$. The matrices $C_{\mu \nu}(J, \alpha)$ represent the
geometry and symmetries of the shell model and are generic in the
sense that they are independent of the actual choice of the residual
interaction. The seemingly trivial Eq.~(\ref{4}) actually contains
very useful information about chaos in the shell model.

To answer the question ``How does the residual interaction cause
chaos?''  in the most general terms, we wish to study the
Hamiltonian~(\ref{4}) for the most general two--body interaction. This
is accomplished by assuming that the elements $v(\alpha)$ of the
two--body interaction are Gaussian--distributed random variables with
mean value zero and a common second moment. The value of that moment
is irrelevant since it determines only the overall scale of the
spectrum. With this assumption, the matrices~(\ref{4}) form an
ensemble of Gaussian--distributed random matrices. This is the
two--body random ensemble (TBRE) of the shell model originally
introduced in the 1970s by French and Wong~\cite{Fre} and by Bohigas
and Flores~\cite{Boh2}. The TBRE is tailored to the shell model and,
therefore, much more realistic than the GOE. The ensemble being
defined in terms of an integration over the Gaussian variables
$v(\alpha)$, statements derived by averaging over the ensemble hold
for all members of the ensemble with the exception of a set of measure
zero with respect to that integration measure.

Within the TBRE, the $v(\alpha)$ play a minor role only: They
determine the specific linear combination of the matrices $C_{\mu
\nu}(J, \alpha)$ which forms the Hamiltonian. If the Hamiltonian
causes chaos for almost all choices of the $v(\alpha)$, that
property must be inherent in the matrices $C_{\mu \nu}(J,
\alpha)$. Therefore, an analytic theory has to address these matrices
as the fundamental building blocks of the TBRE.  Unfortunately, such a
theory is in its infancy as yet. An analytically more accesible
ensemble is provided by the embedded Gaussian orthogonal ensemble
EGOE($k$)~\cite{Mon}. This is defined as a random matrix ensemble of
$m$ Fermions occupying $l$ degenerate single--particle levels with
Gaussian--distributed random $k$-body interaction matrix elements, but
assumes no further symmetries like spin and isospin. For $k=m$, the
EGOE($k$) becomes the GOE, and for $k = 2$, it becomes the analogue of
the TBRE without quantum numbers. For recent reviews, we refer the
reader to Refs.~\cite{Kota,Benet}. Less is kown for the TBRE. In the
sequel, we describe some insights that have been gained over the last
few years.

\subsection{Comparison GOE--TBRE}

The GOE has three important properties: It is invariant under
orthogonal transformations (hence, analytically tractable), it is
universal, and it is ergodic. The TBRE probably does not share any of
these properties. The set of matrices $C_{\mu \nu}(J, \alpha)$ is
fixed. A unitary transformation of all matrices generates another
representation of the ensemble but not another member of the
ensemble. Therefore, the TBRE is not unitarily invariant. It is not
clear how a non--Gaussian distribution of the $v(\alpha)$ would affect
spectral fluctuation properties of the TBRE. In the case of the GOE,
local fluctuation properties and global spectral properties become
separated in the limit $N \to \infty$. That separation is at the root
of universality (local fluctuation properties do not depend on the
form of the distribution of the matrix elements). By definition, the
TBRE is linked to a specific shell; its matrices have finite
dimension. Similarly for ergodicity: For the GOE, that property is
proved by showing that correlation functions vanish with increasing
distance of their energy arguments. The proof uses the limit $N \to
\infty$. This limit does not exist in the TBRE. One is tempted to ask:
Why bother with the TBRE? The answer is: The TBRE is more realistic
than the GOE. Moreover, it might be possible to study the TBRE
analytically for the case of a single $j$--shell and to prove
universality and ergodicity in the limit $j \to \infty$.

Except for the mirror symmetry about the main digonal, every element
of a GOE Hamiltonian matrix stands for an independent random
variable. For $N \gg 1$, the number $N(N+1)/2$ of such variables is
much larger than the matrix dimension $N$. The number $n$ of
independent random variables in the TBRE is typically small in
comparison with the matrix dimension.  For the case of a single
$j$--shell, $n$ grows linearly with $j$ while the typical matrix
dimension grows exponentially with $j$. In the GOE, the analogues of
the matrices $C_{\mu \nu}(J, \alpha)$ exist. These are the $N(N+1)/2$
matrices ${G}_\mu$ which either have a unit element somewhere in the
main diagonal and zeros everywhere else, or have a unit element
somewhere above the main diagonal, its mirror image below, and zeros
everywhere else.  The set $\{G_\mu\}$ forms a complete basis for real
and symmetric matrices. In contradistinction, the matrices $C_{\mu
\nu}(J, \alpha)$ do not form such a complete set. To be sure, every
matrix $C_{\mu \nu}(J, \alpha)$ may be thought of as a linear
combination of the ${G}_\mu $. But the number of matrices $C_{\mu
\nu}(J, \alpha)$ is typically much smaller than $N(N+1)/2$. Therefore,
many other linear combinations of the ${G}_\mu$ exist which are
linearly independent of the $C_{\mu \nu}(J, \alpha)$ and which do not
occur in the TBRE. The TBRE may be negatively defined by constraining
all such linear combinations to be zero.

We are going to present evidence that the matrices $C_{\mu \nu}(J,
\alpha)$ are the agents which cause chaos in the TBRE. In anticipation
of this result, we give a qualitative argument why every matrix
$C_{\mu \nu}(J, \alpha)$ may be thought of as a single representation
of the GOE. In constructing the many--body basis states $| J \mu
\rangle$, we may proceed as follows. We form Slater determinants by
distributing $m$ nucleons over the single--particle states of a major
shell. These are antisymmetric by construction but do not possess good
quantum numbers $J$. States with good $J$ are obtained as linear
combinations of such Slater determinants, a typical state $| J \mu
\rangle$ consisting of a linear combination of {\it many}
determinants. Although the two--body interaction has non--vanishing
matrix elements only between determinants that differ in the
occupation numbers of not more than two single--particle states, that
fact makes sure that the matrices $C_{\mu \nu}(J, \alpha)$ are densely
filled. Moreover, the coefficients in the linear combinations are
determined by angular--momentum algebra, i.e., contain Clebsch--Gordan
coefficients, Racah coefficients, and coefficients of fractional
parentage.  Although individually determined group--theoretically,
these quantities combine to make the coefficients of the linear
combinations almost random numbers. This stochastic aspect of the
shell model has been emphasized by Zelevinsky and collaborators who
employed the term ``geometric chaoticity''~\cite{Zel}.

Thus, the matrices $C_{\mu \nu}(J, \alpha)$ are dense and their
elements are close to being random. Our qualitative argument is
subject to two caveats. First, the dimension of the matrices $C_{\mu
\nu} (J, \alpha)$ is always finite. Second, the matrices $C_{\mu
\nu}(J, \alpha)$ are not dense everywhere and may have a
substructure with blocks that are completely empty. Consider, for
instance, the $sd$--shell and the two--body operator $\alpha_0$ which
scatters a pair of nucleons in states $(d_{3/2}, d_{3/2})$ into states
$(d_{5/2}, d_{5/2})$. That operator has vanishing matrix elements
between all many--body states $| J \mu \rangle$ which are constructed
by filling only the $s_{1/2}$ shell and the $d_{3/2}$ shell with $m$
nucleons. The same statement applies to the matrix $C_{\mu \nu}(J,
\alpha_0)$.

\section{Properties of the TBRE}

In this Section we display several properties of the TBRE which are
either fundamentally different from those of the GOE or have no
counterpart in the latter. The material in this Section is largely
taken from Refs.~\cite{ Pap1,Pap2,Pap3}. The numerical examples are
all obtained in the $sd$--shell or in a single $j$--shell.

\subsection{How the matrices $C_{\mu \nu}(J, \alpha)$ mix the states}

\begin{figure}[h]
\vspace{5 mm}
\centerline{\includegraphics[width=0.4\textwidth]{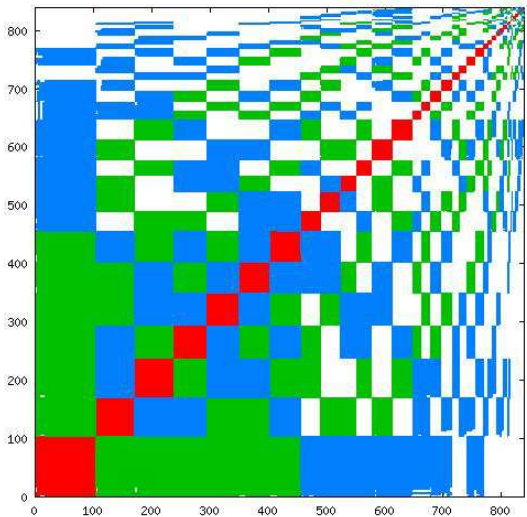}}
\vspace{3 mm}
\caption{(color online)
The block structure of the matrices $C_{\mu \nu}(J, \alpha)$
for $m = 12$ nucleons in the $sd$--shell coupled to total spin $J = 0$
and total isospin $T = 0$. Red: No change of partition. Green: Change
of partition by moving one nucleon. Blue: Change of partition by moving
two nucleons. White: There is no two--body matrix element that connects
the states in question. Taken from Ref.~\cite{Pap2}.}
\label{fig14}
\end{figure}

The $sd$--shell has subshells $s_{1/2}, d_{5/2}, d_{3/2}$. Let $n_i$
with $i = 1,2,3$ denote the occupation numbers of these three
subshells, and $m = \sum_i n_i$ the total number of nucleons in the
$sd$--shell. The many--body basis states $| J \mu \rangle$ can be
ordered in blocks with fixed values of $\{n_1,n_2,n_3\}$. For $m = 12$
there exist 41 such partitions. The two--body matrix elements
$v(\alpha)$ belong to one of three classes: (i) They leave the
partition unchanged so that $\{n_1, n_2,n_3\} \to \{n_1,n_2,n_3\}$ (28
two--body matrix elements); (ii) they change the partition by moving
one nucleon from one subshell to another while leaving the second
nucleon in its subshell so that $\{n_1,n_2,n_3\} \to \{n_1 + 1,n_2 -
1,n_3\}$ etc. (22 two--body matrix elements); (iii) they change the
partition by moving both nucleons from one subshell to another so that
$\{n_1,n_2,n_3\} \to \{n_1 - 1, n_2 - 1, n_3 + 2\}$ cyclic or to
$\{n_1 - 2, n_2 + 1, n_3 + 1\}$ cyclic (13 two--body matrix
elements). The same classification applies to the matrices $C_{\mu
\nu}(J, \alpha)$ since each of these and the corresponding $v(\alpha)$
contain the same two--body operator. Therefore, the $C_{\mu \nu}(J,
\alpha)$ acquire block structure. For 12 nucleons coupled to $J = 0, \
T = 0$ this is shown in Fig.~\ref{fig14} where the 839 states are
ordered according to the sizes of the blocks to which they belong.

The block structure is clearly a consequence of the shell model with
its subshells. It is bound to occur in every major shell. The
structure is completely absent in the GOE where the matrices are
densely filled. It is clear that a single matrix $C_{\mu \nu}(J,
\alpha)$ cannot mix the states completely. Such mixing can be
accomplished only by a linear combination of most or all of the
$C_{\mu \nu}(J, \alpha)$ as in Eq.~(\ref{4}). Put differently, it is
clear that a single or a few non--vanishing matrix elements
$v(\alpha)$ cannot lead to a complete mixing of the basis states,
i.e., to chaos. We expect that we need most or all of the $v(\alpha)$
for that purpose. This observation illustrates the caveat ``with the
exception of a set of measure zero'' used in Subsection~\ref{tbre}:
Putting one or several of the $v(\alpha)$'s identically equal to zero
reduces the set $\{v(\alpha)\}$ to a set of measure zero with respect
to the integration over all the variables $v(\alpha)$. The observation
also sheds light on and reinforces the concept ``geometric
chaoticity''~\cite{Zel}. Indeed, the mixing is both, made possible and
limited, by geometric constraints that result from the coupling of
single-particle states to many-body states with good quantum numbers.

A few numbers may illustrate the way in which the matrices $C_{\mu
\nu}(J, \alpha)$ fill their individual blocks. As a measure for such
filling, we use the number of $C_{\mu \nu}(J, \alpha)$ which
contribute on average to a given matrix element. When the partition is
not changed (red blocks in Figure~\ref{fig14}), that number is $25.2
\pm 4.0$ (out of a total of 28 matrices). Within the red blocks,
mixing is thus very thorough. This statement is supported by another
measure (inverse participation ratios) not shown here. For the green
blocks, the corresponding number is $7.2 \pm 1.4$ (out of a total of
22 matrices) and for the blue blocks, $2.0 \pm 0.9$ (out of a total of
13 matrices). Similar figures are obtained in the case of a single
$j$--shell. The wealth of available numerical data (see, for instance,
Figs.~\ref{fig12} and \ref{fig13}) shows that there is chaos for some
choices of the $v(\alpha)$. This is not an accidental feature related
to a particular choice of the residual interaction.  Indeed, the
numbers just given show that the mixing of states within each of the
diagonal (red) blocks is virtually complete for almost every choice of
the residual interaction. This fact ensures that a few non--diagonal
matrix elements suffice to also mix different diagonal blocks very
efficiently with each other. Thus, the seemingly small numbers cited
for the green and blue blocks are actually quite sufficient to
accomplish complete mixing of all states, even though some (white)
blocks carry zero entries. As a result, we conclude from the structure
of the matrices $C_{\mu \nu}(J, \alpha)$ that chaos is a generic
structural property of the $sd$--shell. Since other major shells are
structurally similar to the $sd$--shell, we conclude that chaos is
generic in the shell model at large.  We formulate this statement with
one proviso: We have neglected the differences in the single--particle
energies pertaining to different subshells. In realistic cases, our
statement applies only if the magnitude of the matrix elements of the
residual interaction is sufficiently large to mix different subshells
sufficiently strongly. That is the case in practice.

While different subshells are mixed quite thoroughly and chaos is
expected to prevail within every major shell, different major shells
do largely retain their identity because the magnitudes of the
$v(\alpha)$ are typically small compared to the spacings of major
shells. This is demonstrated by experimental facts on the behavior of
the neutron strength function or of spectroscopic factors which
unfortunately cannot be presented and explained in the present
framework.

We reiterate the remark that the strong mixing within a major shell
does not preclude the existence of individual states at higher
excitation energies that can be identified as distinct modes of motion
of the nucleus (like the giant--dipole resonances). Deformed nuclei
display collective motion like rotational bands and, in the
ground--state domain, deviations from Wigner--Dyson spectral
fluctuations. This last fact has been extensively discussed in the
literature (see Ref.~\cite{adel} and references to earlier work
therein). At this point, it is not clear how such features can be
reconciled with the view of chaos in nuclei presented in this paper.

\subsection{Information content of nuclear spectra}

GOE spectra do not carry any information content. Spectral
fluctuations in nuclei often agree with GOE predictions. Are such
spectra void of physical information? We answer this provocative
question in quantitative terms~\cite{Pap3} but begin with a
qualitative consideration.

The Hamiltonian~(\ref{4}) contains the matrix elements $v(\alpha)$ as
parameters and is otherwise fixed. The number of these parameters is
typically small compared with the matrix dimension. Therefore, a small
number of data points (for instance, energies of levels with fixed
spin $J$) suffices, in principle, to determine $H_{\mu \nu}(J)$. The
Hamiltonian matrices of the levels with spins $J' \neq J$ are governed
by the ${\it same}$ set of matrix elements $v(\alpha)$ (only the
matrices $C_{\mu \nu}(J', \alpha)$ differ in form and dimension) and
are, therefore, also known once the $v(\alpha)$ are determined.
Hamiltonians describing other nuclei (different mass numbers)
pertaining to the same major shell are likewise governed by the set
$\{v(\alpha)\}$ and, thus, known too.  (Again a caveat is needed: The
set $\{v(\alpha)\}$ which empirically describes the data best may
change with mass number). Thus, the situation is radically different
from that of the GOE because the matrices $C_{\mu \nu}(J, \alpha)$
(the agents of chaos) are fixed by the shell model. A small number of
parameters determines spectra of many $J$ values in many nuclei. TBRE
spectra do carry information.

We turn to the quantitative question: How reliably can this
information be deduced from data? To this end we cast the
Hamiltonian~(\ref{4}) in another form. We seek a decomposition of the
Hamiltonian into a sum of matrices that can be ordered with respect to
their ``relevance'' or ``importance''. The decomposition~(\ref{4}) is
not useful for this purpose: The matrices $C_{\mu\nu}(J,\alpha)$
exhibit some degree of linear independence but a stronger metric
concept is called for. For this purpose, we introduce the canonical
scalar product for matrices, and switch from the matrices
$C_{\mu\nu}(J,\alpha)$ to new matrices $B_{\mu\nu}(J,\alpha)$ that
form an orthonormal basis set. With $d(J)$ the dimension of the
Hilbert space spanned by the state vectors $| J \mu \rangle$, we
define the real and symmetric overlap matrices
\be
S_{\alpha \beta}(J) = d^{-1}(J) \ {\rm Trace} [ C(J, \alpha) C(J,
\beta) ] \ .
\label{5}
\ee
These are diagonalized by orthogonal transformations ${\cal O}(J)$,
\be
\{ {\cal O}(J) S(J) [ {\cal O}(J) ]^T \}_{\alpha \beta} = s^2_\alpha(J)
\delta_{\alpha \beta}
\label{6}
\ee 
where $T$ denotes the transpose. The matrices $S_{\alpha \beta}(J)$
are positive semi-definite so that $s^2_\alpha(J) \geq 0$. The
eigenvalues are arranged by decreasing magnitude, $s^2_1(J) \geq
s^2_2(J) \geq \ldots \geq 0$. The real roots $s_\alpha(J)$ of the
$s^2_\alpha(J)$'s are chosen positive or zero. If one or several
eigenvalues vanish, we conclude from Eqs.~(\ref{6}) and (\ref{5}) that
there exist one or several linear combinations of the matrices $C(J,
\alpha)$ that vanish identically. For the $a_1(J)$ non--vanishing
eigenvalues $s^2_\alpha(J)$, we define
\be
B_{\mu \nu}(J, \alpha) = \frac{1}{s_\alpha(J)} \sum_\beta
{\cal O}_{\alpha \beta} C_{\mu \nu}(J, \beta) \ , \ \alpha = 1, \ldots,
a_1(J) \ .
\label{7}
\ee
Except for possible degeneracies among the eigenvalues
$s^2_\alpha(J)$, the matrices $B_{\mu \nu}(J, \alpha)$ are defined
uniquely. By construction, these matrices are orthonormal with respect
to the trace,
\be
d^{-1}(J) \ {\rm Trace} [ B(J, \alpha) B(J, \beta) ] = \delta_{\alpha
\beta} \ .
\label{8}
\ee
Written in terms of the matrices $B(J, \alpha)$, the Hamiltonian takes
the form
\be 
H_{\mu \nu}(J) = \sum_{\alpha = 1}^{a_1} w(J, \alpha) s_\alpha(J)
B_{\mu \nu}(J, \alpha) \ . 
\label{9}
\ee
We note that the eigenvalues $s^2_\alpha(J)$ indicate the weight and
relevance of the corresponding basis matrix $B(J,\alpha)$ in the
construction of the Hamiltonian~(\ref{9}). This weight has purely
geometric origin. The dynamical weight $w(J, \alpha)$ stems from the
two-body interaction and is given by
\be
w(J, \alpha) = \sum_\beta {\cal O}_{\alpha \beta} v(\beta) \ , \ \alpha
= 1, 2, \ldots, a_1 \ . 
\label{10}
\ee
The $w(J, \alpha)$ have the same distribution as the $v(\alpha)$: They
are Gaussian-distributed random variables with mean values zero and a
common second moment. The form~(\ref{9}) of the Hamiltonian with the
definitions~(\ref{7}) and (\ref{10}) is completely equivalent to the
original expression~(\ref{4}). It allows us to quantify the
information content of nuclear spectra. If for some value of $J$ one
or several eigenvalues $s^2_\alpha(J)$ vanish, the number $a_1(J)$ of
variables $w(J, \alpha)$ is smaller than the number of elements
$v(\alpha)$. For that value of $J$ it is then impossible to determine
more than $a_1(J)$ linear combinations of the $v(\alpha)$ from
data. We shall see that such vanishing eigenvalues always
occur. Moreover, if some eigenvalues are significantly smaller than
the leading ones, then Eq.~(\ref{9}) shows that their influence on the
spectrum is small (we recall that the matrices $B(J, \alpha)$ obey
Eq.~(\ref{8})), and it will be difficult to deduce their values from
data. We conclude that it is important to determine the eigenvalues
$s^2_\alpha(J)$. That can be done without prior knowledge of the form
of the residual interaction (i.e., of the values of the $v(\alpha)$)
since the eigenvalues are determined by the matrices $C_{\mu \nu}(J,
\alpha)$, i.e., by the shell model itself.

\begin{figure}[h]
\vspace{5 mm}
\centerline{\includegraphics[width=0.4\textwidth]{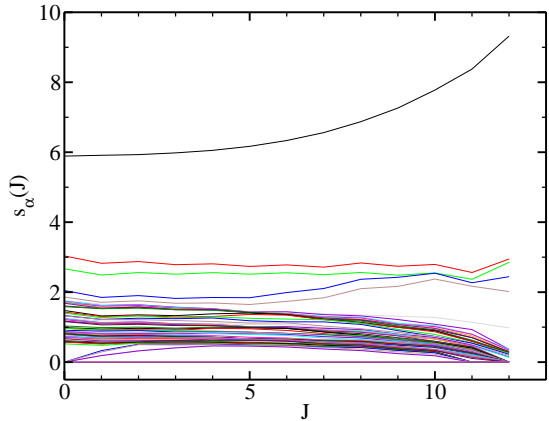}}
\vspace{3 mm}
\caption{Roots $s_\alpha(J)$ of the eigenvalues $s^2_\alpha(J)$
defined in Eq.~(\ref{6}) versus total spin $J$ for the $T = 0$ states
in $^{24}$Mg. Taken from Ref.~\cite{Pap3}.}
\label{fig15}
\end{figure}

In Refs.~\cite{Pap2,Pap3}, the distribution of the roots
$s_{\alpha}(J)$ of the eigenvalues was given for a single $j$--shell
with $j = 19/2$ and for some nuclei in the $sd$--shell. All these
distributions are very similar. By way of example, we reproduce in
Fig.~\ref{fig15} the graph for the states with isospin zero in the
nucleus $^{24}$Mg.

We note that all roots change smoothly with $J$. The largest,
$s_1(J)$, is significantly larger than all others. For the case of the
single $j$--shell, it can be shown analytically that the corresponding
linear combination $B_{\mu \nu}(J, 1)$ of the matrices $C_{\mu \nu}(J,
\alpha)$ is approximately equal to the matrix representation of the
monopole operator which in turn is approximately equal to the unit
matrix. The numerical results indicate that these statements are
generic; they hold also for $^{24}$Mg. Hence, $s_1(J)$ determines
essentially the centroid of the spectrum of states with spin $J$ and
has little influence on spectral fluctuations. The latter are largely
determined by the remaining non--vanishing eigenvalues and associated
matrices $B_{\mu \nu}(J, \alpha)$. Because of the orthogonality
relation~(\ref{8}), the latter are almost traceless. 

The root $s_1(J)$ is followed in magnitude by four eigenvalues that
are distinct from the remaining set. We observe that some eigenvalues
in that set vanish for large values of $J$ (this is because the matrix
dimension shrinks with increasing $J$ and eventually becomes too small
to support a large number of eigenvalues) and that at least one
eigenvalue vanishes for all values of $J$. This is because the
operator ${\vec J}^2$ for total spin $J$ is a two-body operator. Thus,
the matrix representation of the term ${\vec J}^2 - J(J+1)$ can be
written as a linear combination of the matrices $C_{\mu \nu}(J,
\alpha)$. This linear combination vanishes identically.

Figure~\ref{fig15} shows that it is difficult to determine those $w(J,
\alpha)$'s from data which belong to the smallest eigenvalues, and that
the problem cannot be solved by combining data for different spin values.
Practitioners of the shell model avoid the difficulty by combining a fit
to data with ${\it ab \ initio}$ values of the $v(\alpha)$ obtained from
many--body theory, see, for instance, Refs~\cite{Bro,Hon}.

\subsection{Preponderance of ground states with spin zero}

In 1998, Johnson, Bertsch, and Dean~\cite{Joh} reported that in
even--even nuclei (nuclei with even proton and neutron numbers), the
TBRE yields ground states with spin zero much more frequently than
corresponds to the fraction of spin-zero states in the model
space. Subsequent work~\cite{Bij,Jac} showed that similar regularities
exist in bosonic and electronic many--body systems. The phenomenon has
received intense attention, and we refer the reader to the
reviews~\cite{Zel1,Zhao1}. The problem has been understood
quantitatively for bosonic models of the nucleus within mean-field
calculations \cite{Bij01}. For the nuclear shell model, two approaches
yield good agreement with numerical simulations. Zhao {\it et
al.}~\cite{Zhao2} devised an algorithm that accurately predicts the
fraction of ground states with a given spin. The other approach
considers fluctuations of and correlations between the $J$-dependent
spectral widths and also leads to semi-quantitative predictions. It
was put forward in Ref.~\cite{Pap1}, and is described in what follows.

Key to the understanding of the phenomenon is the observation that the
spectral widths $\sigma_J$ of states with different $J$ values are
correlated. The spectral widths are defined by
\be
\sigma^2_J = d^{-1}(J) \ {\rm Trace} [ H^2(J) ] \ .
\label{11}
\ee
From Eqs.~(\ref{9}) and (\ref{8}), we have
\be
\sigma^2_J = \sum_{\alpha = 1}^{a_1} w^2(J, \alpha) s^2_\alpha(J) \ .
\label{12}
\ee
Spectral widths pertaining to different $J$ depend upon the same
random variables $v(\alpha)$ (the $w(J, \alpha)$ are linear functions
of the $v(\alpha)$, see Eq.~(\ref{10})) and are, therefore, correlated.

Before using this fact, we relate the spectral width $\sigma_J$ to the
position of the lowest (or highest) state with spin $J$, more
precisely, to the distance $R_J$ of the lowest (or highest) level in
the spectrum from the origin (the spectral radius). (We do not
distinguish the lowest from the highest state since the $v(\alpha)$
have random signs.) The spectral radius is expected to be a random
variable. We ask for the probability $p_j$ with which, for a given
value of $J$, $R_J$ takes maximum value. The probability of finding a
ground state with spin $J$ is given by $p_J$.

\begin{figure}[h]
\vspace{5 mm}
\centerline{\includegraphics[width=0.4\textwidth]{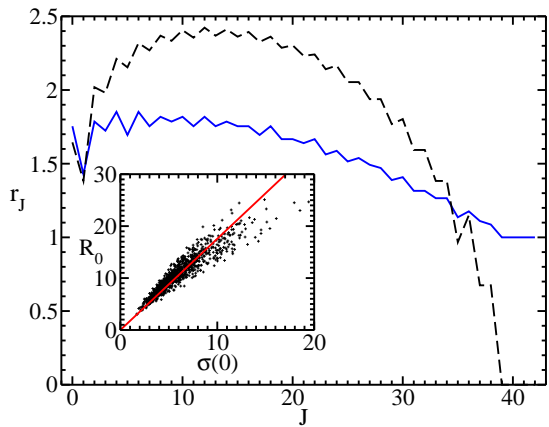}}
\vspace{3 mm}
\caption{(color online)
The scaling factor $r_J$ of Eq.~(\ref{13}) versus $J$ for six
fermions in the $j = 19/2$ shell (solid line). 
Inset: Linear fit to $r_J$ for the
states with $J = 0$. The dots correspond to 900 realizations of the
ensemble. Taken from Ref.~\cite{Pap1}.}
\label{fig16}
\end{figure}

We define the scaling factor $r_J$ by writing
\be
R_J = r_j \sigma_J \ . 
\label{13}
\ee
Eq.~(\ref{13}) is useful in the present context because it turns out
that (in contrast to $\sigma_J$) $r_J$ hardly fluctuates with the
$v(\alpha)$. For the case of 6 fermions in the $j = 19/2$ shell and
for the states with spin $J = 0$, this is shown in the inset of
Fig.~\ref{fig16}. Similarly constant (non--fluctuating) behavior of
$r_J$ was found for other cases (different spin values, different
number of nucleons, nuclei in the $sd$--shell). Therefore, the
fluctuations and correlations of the widths $\sigma_J$ directly affect
those of the spectral radii $R_J$. The linear fits to $r_J$, obtained
as shown in the inset, are used for the plot of $r_J$ versus $J$ shown
in Fig.~\ref{fig16} (and similarly for other cases). The overall
monotonic decrease of $r_J$ with $J$ reflects the overall monotonic
decrease of the dimensions $d(J)$ with $J$. Indeed, average
shell--model spectra are known to have approximately Gaussian
shape. It is then intuitively obvious that $r_J$ grows with
$d(J)$. The decrease of $r_J$ with increasing $J$ shown in the figure
enhances the chance for states with small spins $J$ to form the ground
state. The same is true of the odd--even staggering of $r_J$ (which is
due to the same reason and, everything else being equal, gives, for
instance, preference to spin zero over spin 1).

\begin{figure}[h]
\vspace{5 mm}
\centerline{\includegraphics[width=0.4\textwidth]{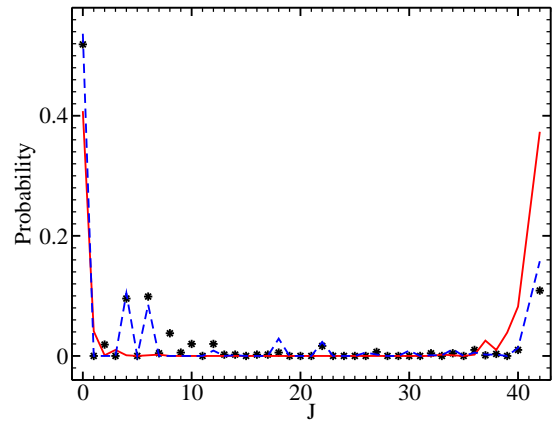}}
\vspace{3 mm}
\caption{(color online) Six fermions in the $j = 19/2$ shell with 900
realizations of the TBRE. Solid red line: Numerical probabilities for
$\sigma_J$ to have maximum value. Dashed blue line: Numerical
probabilities for $r_J \sigma_J$ to have maximum value.  Dots:
Numerical probabilities for state with spin $J$ to be the ground
state. Taken from Ref.~\cite{Pap1}.}
\label{fig17}
\end{figure}

It remains to study the probabilities that the spectral widths
$\sigma_J$ attain maximum values. For the case displayed in
Fig.~\ref{fig16}, these are shown as red lines in
Fig.~\ref{fig17}. The states with lowest and highest spins have the
largest probabilities. Here the correlations play a role: For some
choice of the $v(\alpha)$, the spectral width for some other spin
value may take an unusually large value but so do, at the same time
and for most cases, $\sigma_0$ and $\sigma_{42}$. The blue line in
Fig.~\ref{fig17} gives the probability that the product $r_J \sigma_J$
attains maximum value. Because of the effect shown in
Fig.~\ref{fig16}, the highest spin value is suppressed, and spin zero
wins. The result must be compared with the actual probabilities (dots)
that a state with spin $J$ forms the ground state. The agreement,
although not perfect, shows that the explanation accounts for the main
features of the phenomenon.

\begin{figure}[h]
\vspace{5 mm}
\centerline{\includegraphics[width=0.4\textwidth]{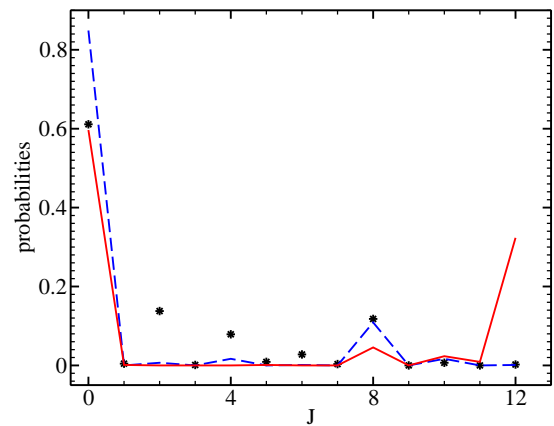}}
\vspace{3 mm}
\caption{(color online)
Same as Fig.~\ref{fig17} for the $T = 0$ states in $^{24}$Mg.
Taken from Ref.~\cite{Pap3}.}
\label{fig18}
\end{figure}

Similar results were obtained for eight Fermions in the $j = 19/2$
shell, for $^{20}$Ne, and for $^{24}$Mg. Figure~\ref{fig18} shows the
case of states with $T = 0$ in $^{24}$Mg.

\subsection{Correlations between spectra carrying different quantum
numbers}

In the TBRE the preponderance of ground states with spin zero is, to a
large extent, due to the correlations between spectral widths
$\sigma_J$ pertaining to different values of $J$. The correlations are
caused by the fact that all widths $\sigma_J$ depend on the same
random variables $v(\alpha)$. Eq.~(\ref{4}) shows that the same
statement holds for the Hamiltonians $H_{\mu \nu}(J)$ pertaining to
different spin values and, more generally, for the Hamiltonians
pertaining to nuclei belonging to the same major shell.

From the point of view of the shell model, these statements are not
terribly surprising. A change of the residual interaction causes
simultaneous changes in the spectra of all nuclei belonging to the
same major shell. In the framework of a statistical description, this
fact is tantamount to the existence of spectral correlations. Still,
it is remarkable that such correlations have never been addressed in
the framework of RMT until quite recently. It is, in fact, not obvious
how to model such correlations in canonical RMT. The standard approach
consists of assuming that correlations do not exist. This assumption
is usually stated explicitly in the statistical theory of nuclear
reactions. Let $S_{a b}(E, J)$ be an element of the statistical
scattering matrix for scattering from channel $a$ into channel $b$ at
energy $E$ and total spin $J$. It is assumed that elements pertaining
to different $J$ values are uncorrelated. This assumption implies the
symmetry of compound--nucleus scattering cross sections about $90$
degrees scattering angle in the center--of--mass system.

It is obviously of interest to test these standard assumptions of RMT
using the TBRE. While tests for correlations of the statistical
scattering matrix have apparently not yet been performed, tests of
spectral correlations for levels both in the same nucleus carrying
different quantum numbers, and in different nuclei do
exist~\cite{Pap3}. The level density $\rho(E, J)$ for levels with spin
$J$ and energies $E_\mu(J)$ (where $\mu$ is a running index) is given
by
\be
\rho(E, J) = \sum_\mu \delta( E - E_\mu(J)) \ .
\label{14}
\ee
As a measure of spectral correlations, we use the correlator
\ba
C(E_1, J_1; E_2, J_2) &=& \overline{\rho(E_1, J_1) \rho(E_2, J_2)}
\nonumber \\
&& \ \  - \overline{\rho(E_1, J_1)} \ \overline{\rho(E_2, J_2)} \ .
\label{15}
\ea

\begin{widetext}
\begin{figure*}[t]
\centerline{
\includegraphics[width=0.28\textwidth]{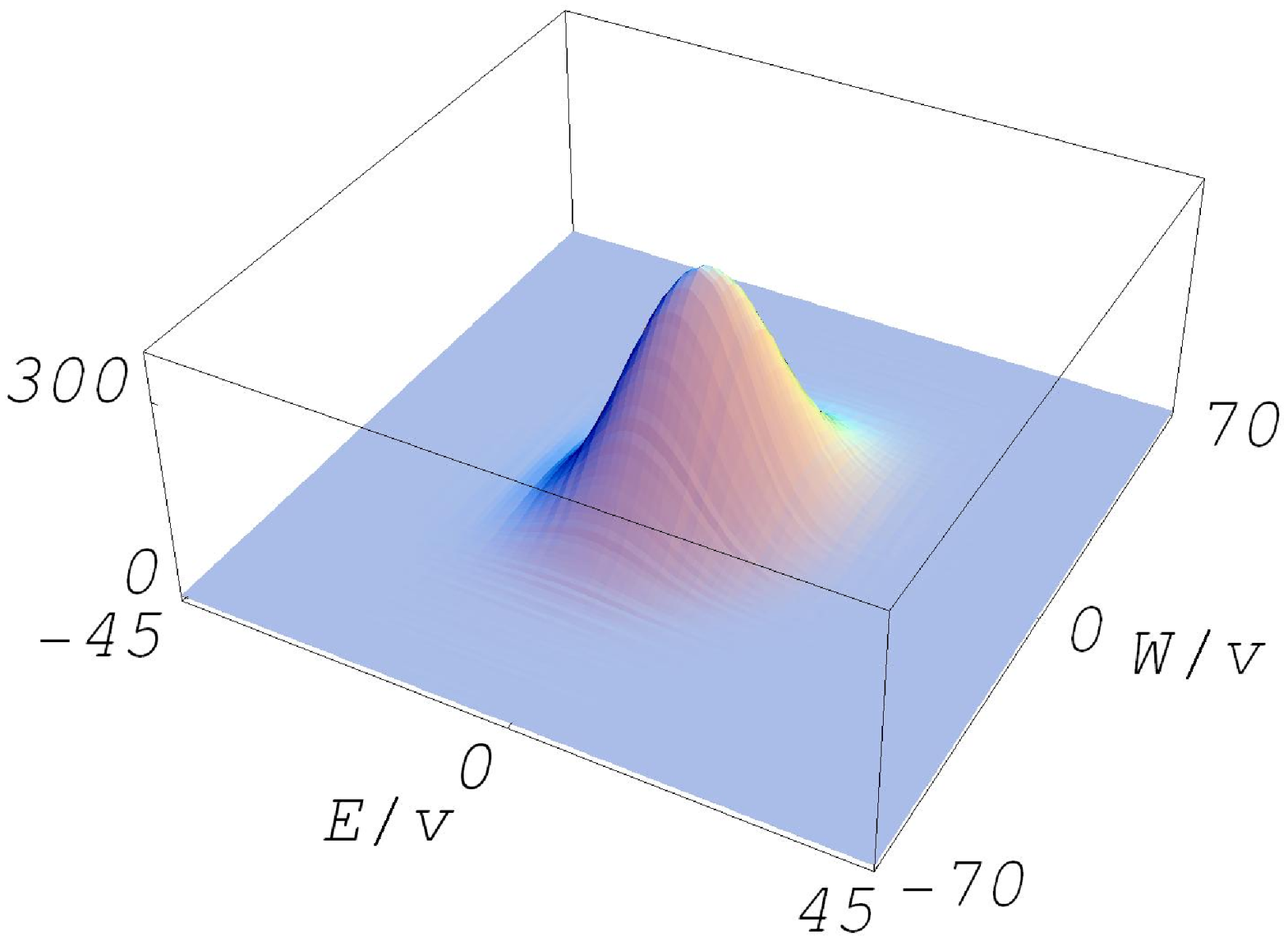}
\includegraphics[width=0.28\textwidth]{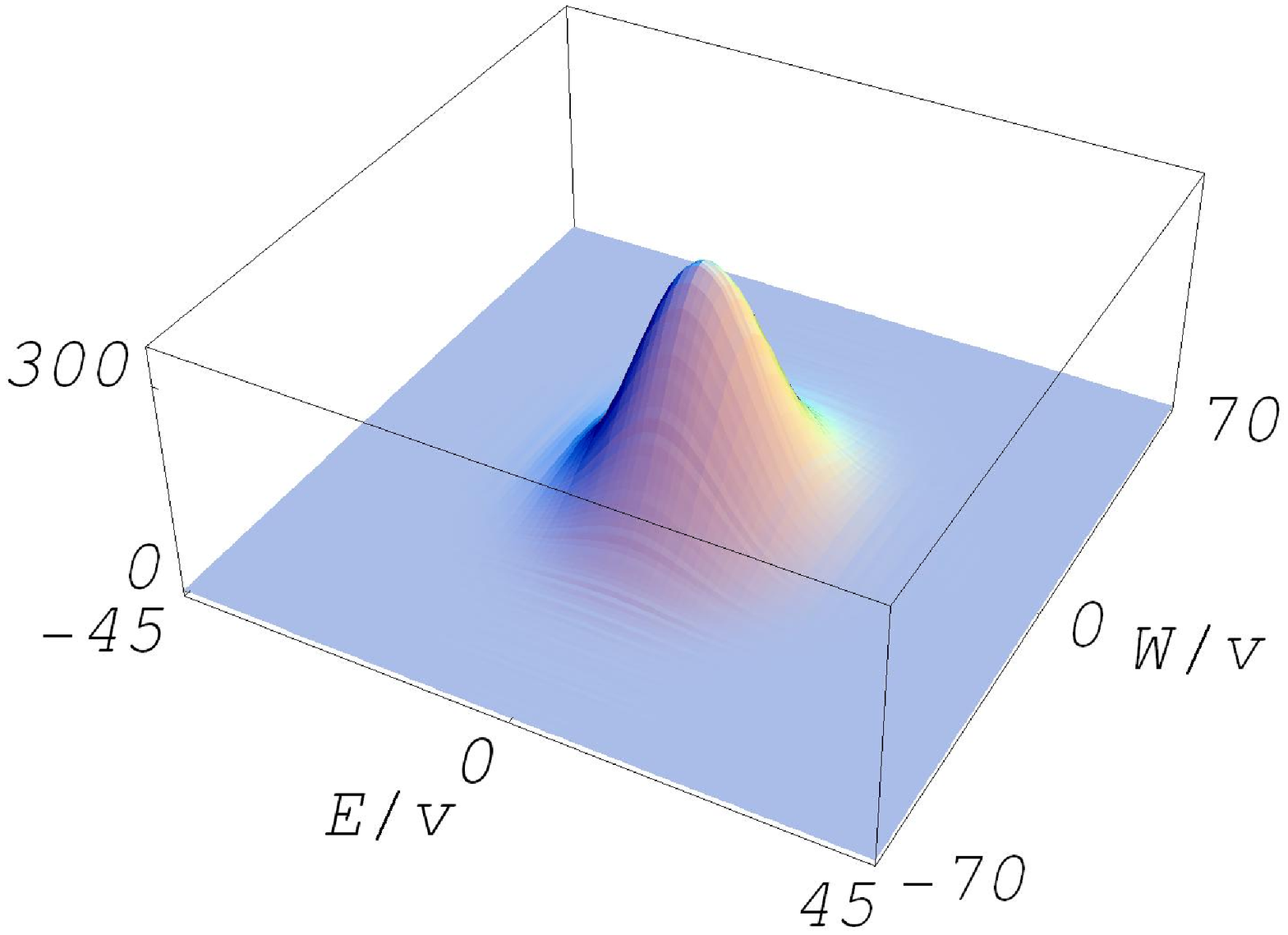}
\includegraphics[width=0.28\textwidth]{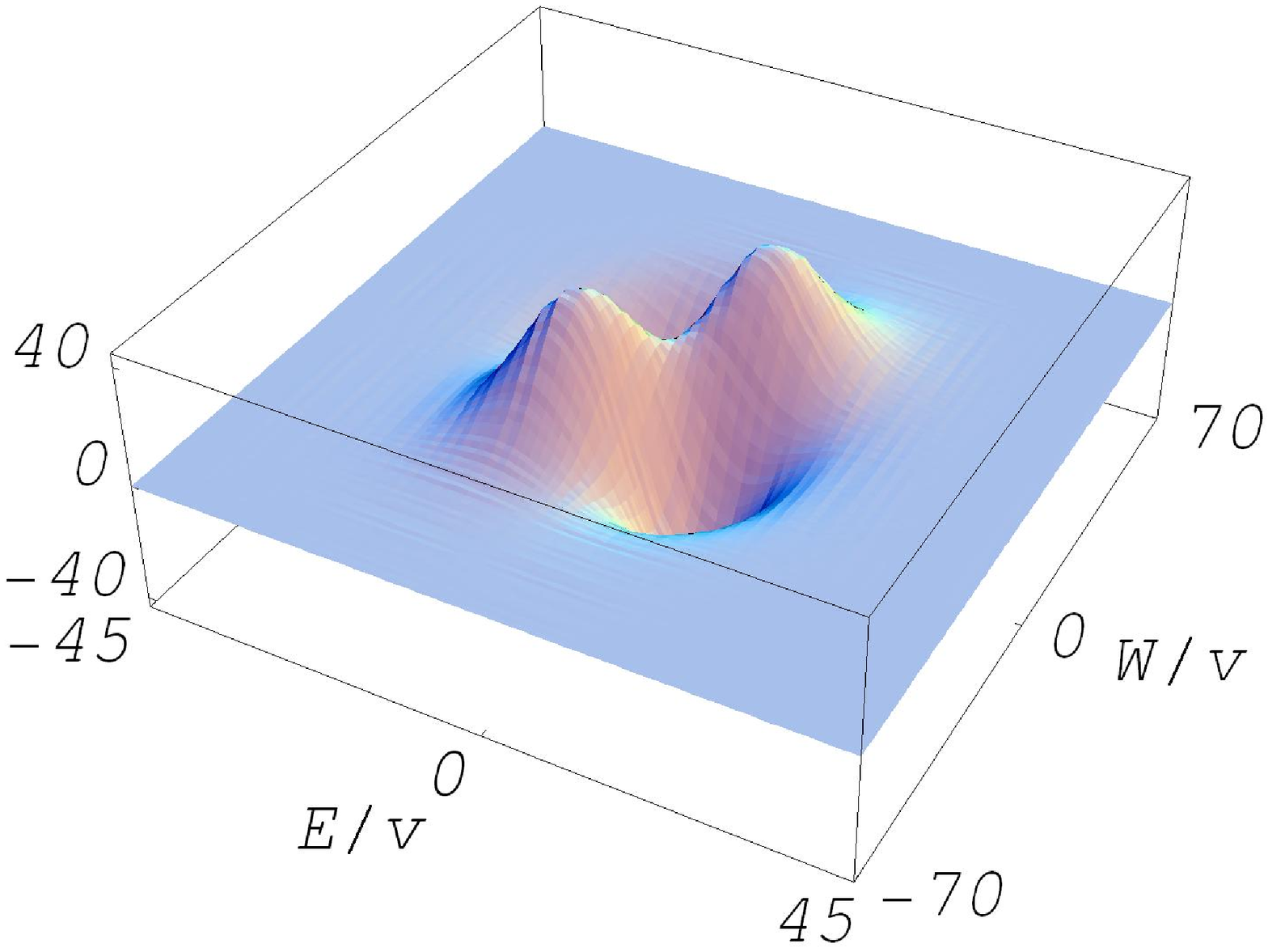}}
\caption{Correlations between the level densities for states with
spin $0$ and spin $2$ (both with $T = 0$) in $^{24}$Mg. Left panel:
Mean value of the product of the two level densities. Middle panel:
Product of the mean level densities. Right panel: The correlator of
Eq.~(\ref{15}). (From 400 realizations of the ensemble). Taken from
Ref.~\cite{Pap3}.}
\label{fig19}
\end{figure*}
\end{widetext}

In Fig.~\ref{fig19} we show the two terms on the right--hand side of
Eq.~(\ref{15}) and the resulting correlator versus the energies $E_1,
E_2$ of the two sets of spin states for the $J = 0, T = 0$ and the $J
= 2, T = 0$ states in $^{24}$Mg. The correlator has a maximum value
of about $13$ percent of the mean value of the product of the level
densities. 

\begin{widetext}
\begin{figure*}[bh]
\centerline{
\includegraphics[width=0.28\textwidth]{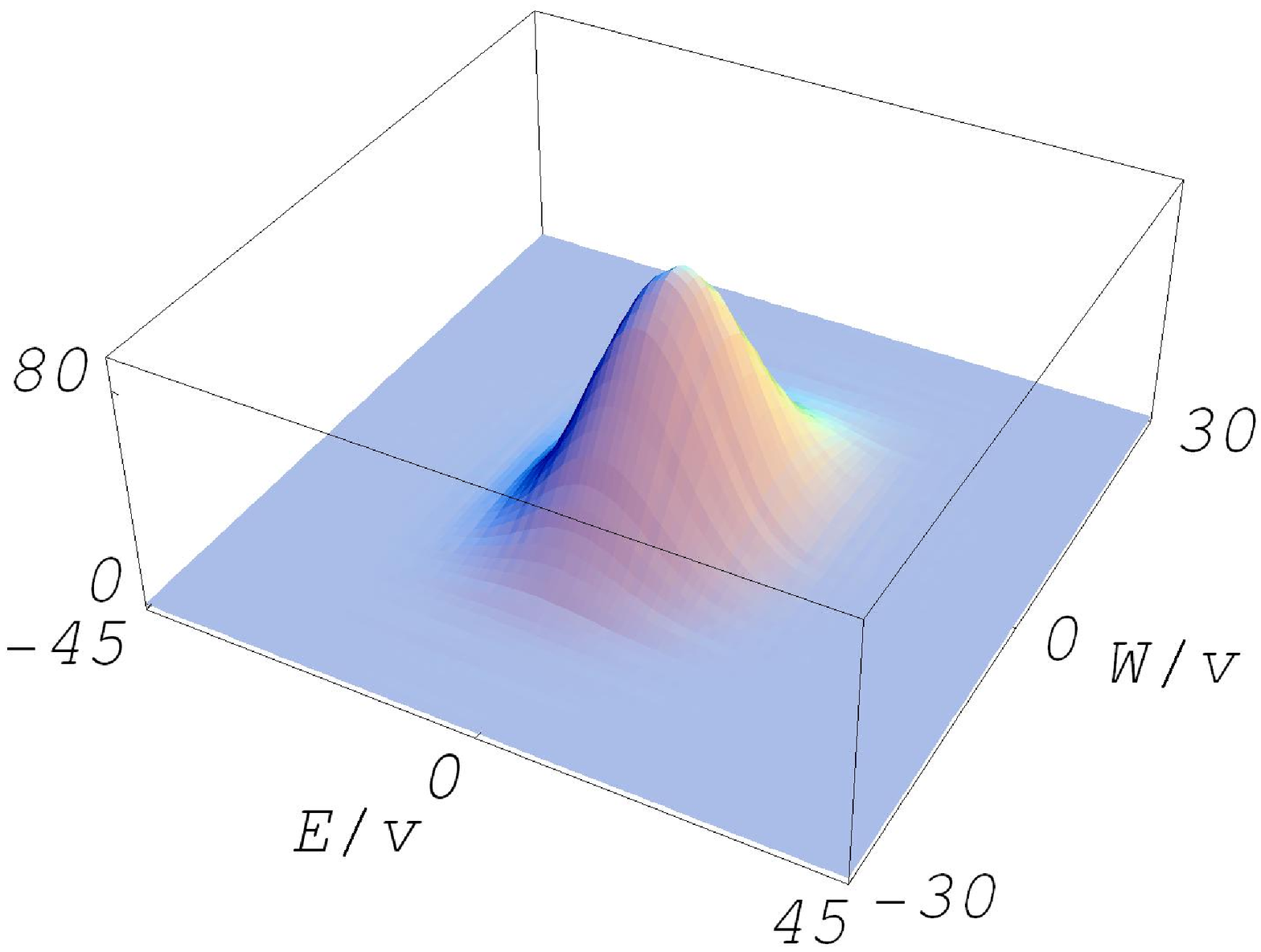}
\includegraphics[width=0.28\textwidth]{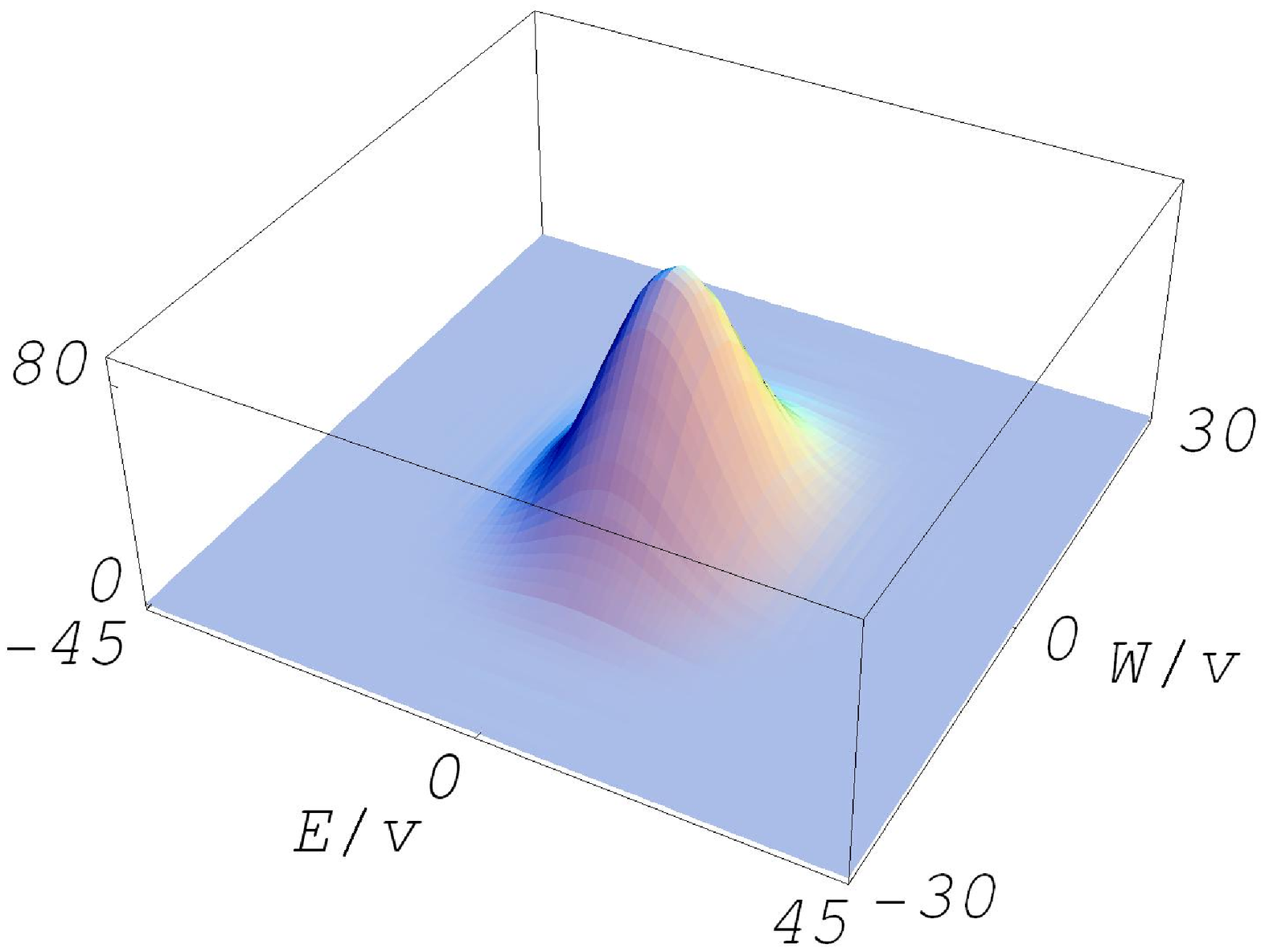}
\includegraphics[width=0.28\textwidth]{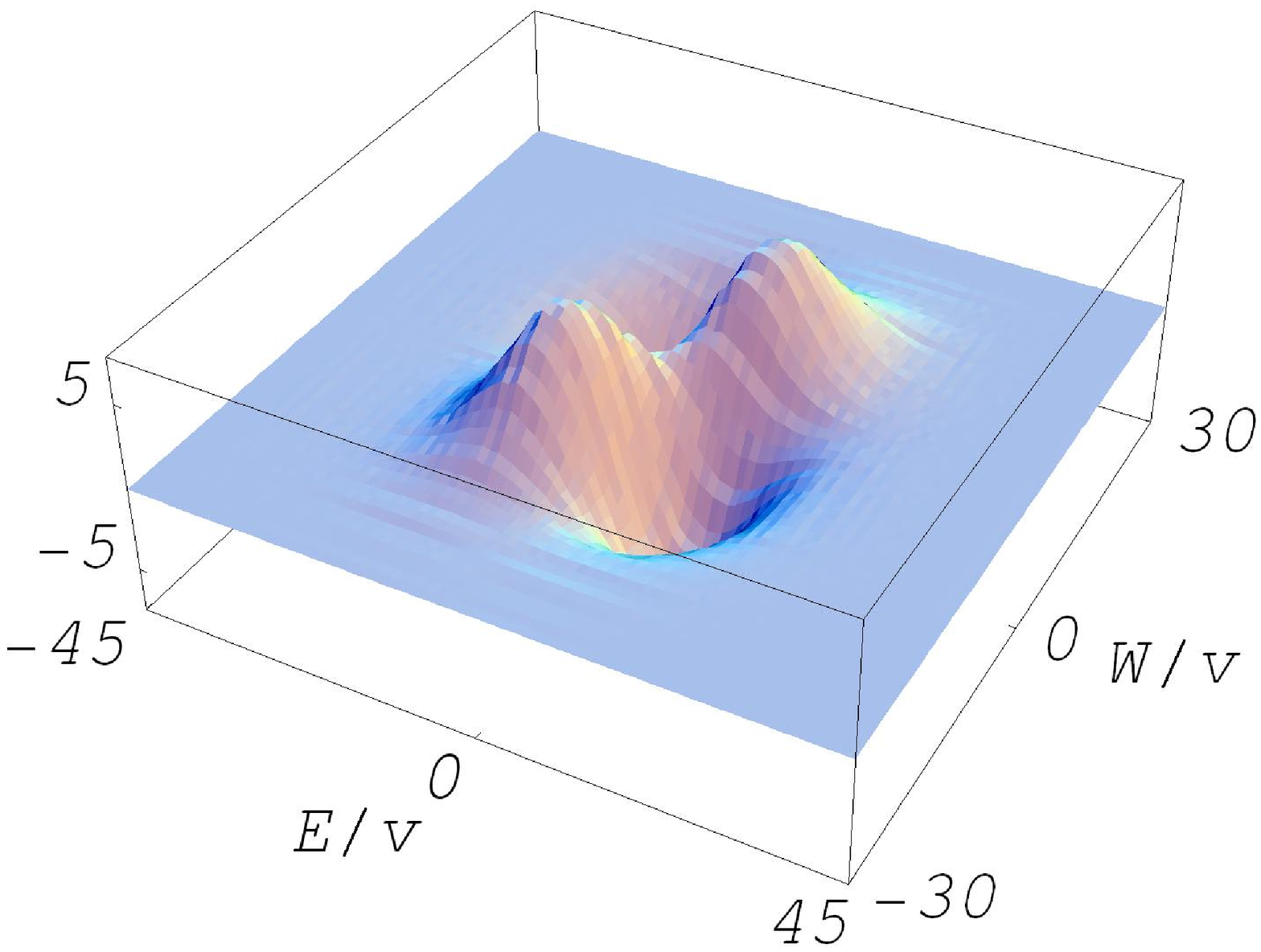}}
\caption{Same as Fig.~\ref{fig19} except for the correlations
between the level densities for states with spin $0$ and isospin $0$
in $^{20}$Ne and in $^{24}$Mg. Left panel: Mean value of the product
of the two level densities. Middle panel: Product of the mean level
densities.  Right panel: The correlator of Eq.~(\ref{15}). (From 400
realizations of the ensemble). Taken from Ref.~\cite{Pap3}.}
\label{fig20}
\end{figure*}
\end{widetext}

In Fig.~\ref{fig20} we show the same quantities for the
$J = 0, T = 0$ states in $^{22}$Ne and in $^{24}$Mg. Here the
correlator has a maximum value of $6$ percent. In both cases, the
existence of spectral correlations is definitely established.

\begin{widetext}
\begin{figure*}[th]
\centerline{
\includegraphics[width=0.28\textwidth]{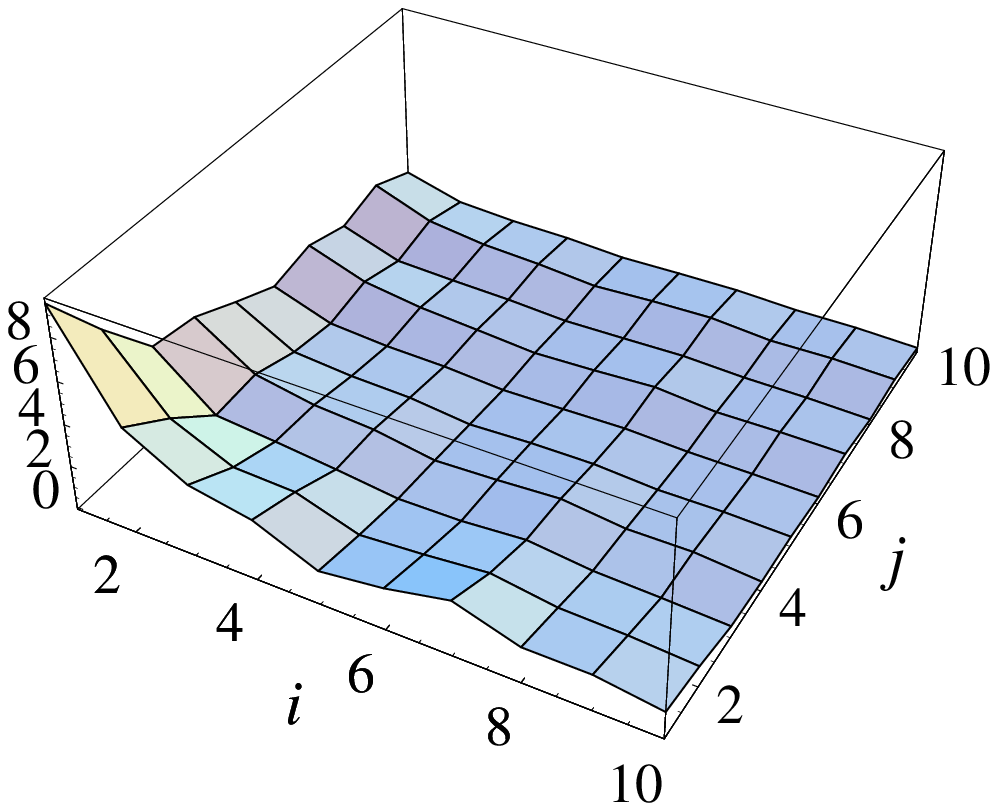}
\includegraphics[width=0.28\textwidth]{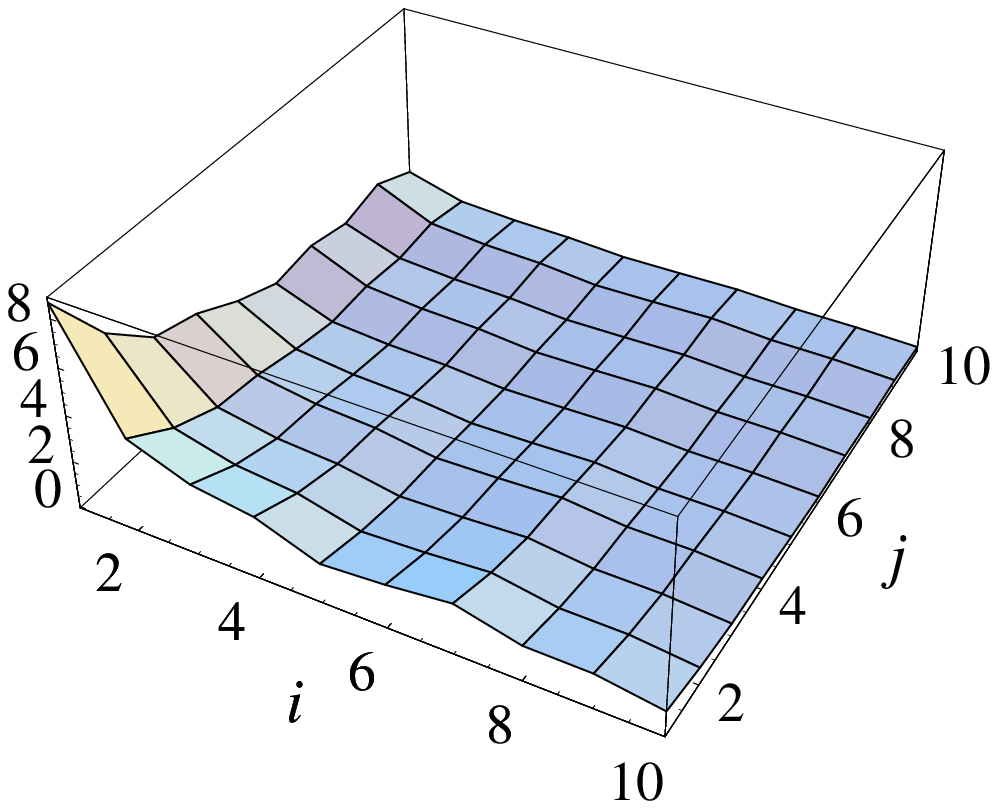}
\includegraphics[width=0.28\textwidth]{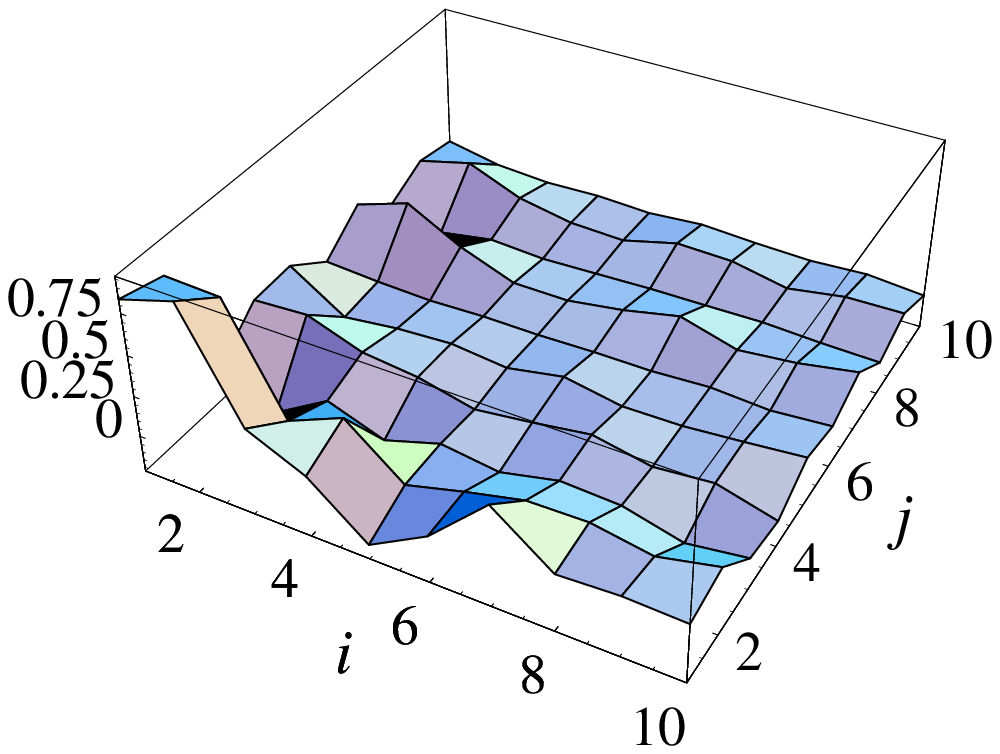}}
\caption{Same as Fig.~\ref{fig19} except for the correlations
between pairs of nearest--neighbor spacings of low--lying levels with
different spins obtained by averaging over a number of nuclei in the
$sd$--shell (see text). The indices $i$ and $j$ label the spacings
consecutively starting from the lowest state. Left panel: Mean value
of the product of the two spacings. Middle panel: Product of the mean
values of the two spacings. Right panel: The correlator. Taken from
Ref.~\cite{Pap3}.}
\label{fig21}
\end{figure*}
\end{widetext}

The correlators shown in Figures~\ref{fig19} and \ref{fig20} are
obtained by averaging over the ensemble and are, therefore, entirely
theoretical. Do they have any correspondence in reality (where we deal
with a single Hamiltonian rather than an ensemble)? Presently, this
question cannot be answered using experimental data -- the data set is
not sufficient. However, shell--model calculations can serve as a
substitute. Realistic single--particle energies and an optimized
two--body interaction (the ``Brown--Wildenthal
interaction''~\cite{Bro}) were used to calculate the energies of
low--lying states with spins $J = 0$ and $J = 2$ ($J = 1/2$ and $J =
5/2$) in a number of even--even (odd--even, respectively) nuclei in
the $sd$--shell~\cite{Pap3}. The correlations between
nearest--neighbor level spacings of the lowest few states with
different spins were evaluated as in Eq.~(\ref{15}), the ensemble
average being replaced by the running average over the set of nuclei
just mentioned. Figure~\ref{fig21} shows the result (displayed in the
same form as in Figs.~\ref{fig19} and \ref{fig20}). Again, the
existence of correlations is established. We observe that these are
particularly pronounced for the lowest levels where they amount to
about $10$ percent.

Our analysis establishes the existence of correlations between spectra
carrying different quantum numbers. All examples were taken from the
$sd$--shell, however, where the dimensions of the Hamiltonian matrices
are rather small (typically $10^2$ to $10^3$). The maxima of the
correlators are in the $10$ percent range. It is not clear how these
results carry over to other major shells where both the matrix
dimensions and the number of independent two--body matrix elements are
much bigger. The statistical theory of nuclear reactions is mainly
used in such shells. Therefore, definitive conclusions must wait for a
further analysis.

\section{Summary and Conclusions}

In the first part of this paper, we have introduced random matrices
and the concept of (quantum) chaos. Both are linked by the
Bohigas--Giannoni--Schmit conjecture. Random--matrix theory predicts
fluctuation properties of spectra in a parameter--free fashion. In
many cases, the fluctuation properties of nuclear spectra agree with
these predictions: The nuclear dynamics is (partly) chaotic. To relate
this observation to known dynamical features of spherical nuclei, we
have, in the second part, described the nuclear shell model (a
mean--field theory with residual interactions). We have pointed out
that shell--model spectra also show chaos. To explain this
observation, we have defined the two--body random ensemble (TBRE) as
the generic Gaussian random--matrix ensemble of the shell model. In
part three, we have displayed a number of properties of that ensemble,
mainly using the $sd$--shell. We have focused attention on those
properties in which the TBRE differs from standard RMT: (i) The TBRE
generically produces chaos. Thus, chaos in nuclei is not due to a
particular feature of the residual interaction.  The mechanism which
mixes the basis states of Hilbert space differs significantly from
that of canonical RMT. Chaos is due to the ``building block'' matrices
$C_{\mu \nu}(J, \alpha)$. In each major shell, these matrices are
fixed once and for all. They are determined entirely by intrinsic
properties of the shell model and are independent of any particular
residual interaction. (ii) The spectra of matrices drawn at random
from a canonical ensemble of RMT do not carry any information
content. Because of the role played by the fixed matrices $C_{\mu
\nu}(J, \alpha)$, this is not so for the TBRE. On the contrary, it is
possible to extract relevant physical information from nuclear
spectra. We have analyzed the degree to which this can be done. (iii)
The preponderance of ground states with spin zero in the TBRE is, to a
large extent, due to correlations between widths of spectra with
different spin values. (iv) Correlations exist more generally also
between spectra carrying different quantum numbers and in different
nuclei belonging to the same major shell. The implications of this
last statement have not yet been fully explored.

The examples used to illustrate or prove these assertions were taken
from the $sd$--shell or from a single $j$--shell. The line of
reasoning suggests very strongly, however, that the spectral
fluctuation properties of the TBRE agree with RMT predictions for all
major shells.  We have confined ourselves to a two--body residual
interaction and to the TBRE. It is obvious that the inclusion of
three--body forces would enhance the tendency of the system towards
chaotic dynamics.

Throughout most of the paper, we have assumed that the
single--particle energies of subshells belonging to a major shell are
degenerate. Within this approximation, we have shown that quantum
chaos is a generic feature of the nuclear shell model. This means that
most choices of the residual interaction (but not all) will result in
spectra showing spectral fluctuations that agree with RMT
predictions. Actually, the neglect of the differences of
single--particle energies of subshells is not completely
justified. The magnitude of a typical two--body matrix element which
mixes different subshells is comparable with the difference of the
corresponding single--particle energies. The resulting incomplete
mixing of many--body states within a major shell reveals itself in
slight deviations from RMT predictions~\cite{Zel}.  Moreover, the
residual interaction is far too weak to thoroughly mix adjacent major
shells. This is why shell structure remains the hallmark of nuclear
physics, in spite of quantum chaos. It is in that sense that spherical
nuclei display only partial quantum chaos.

Low--lying states in spherical nuclei often display regular features.
This statement is not at variance with our conclusions. Indeed, chaos
manifests itself mainly in the fluctuation properties of spectra.
These are defined in terms of statistical measures tests of which
require a large number of levels with identical quantum numbers. The
regular features just mentioned refer to properties of a small number
of levels in the ground--state domain which carry different quantum
numbers. The properties of these levels may be particularly sensitive
to a specific component of the two--body interaction. Modelling the
entire residual interaction in terms of that component may give rise
to regular motion.

Our discussion has been confined to spherical nuclei. Nuclei with mass
numbers lying in the middle of large major shells (like the
rare--earth nuclei) cannot be successfully described in terms of the
spherical shell model. The dimensions of the many--body Hilbert spaces
are too large.  One uses instead collective models with a small number
of degrees of freedom. The connection between these and the spherical
shell model is not firmly established. However, quantum chaos is
prevalent in these nuclei, too, with the exception of cases of
distinct symmetries~\cite{Whe}.

There are several open questions and directions for future research.
(i) We are still lacking a deeper analytical understanding of the TBRE
and its fluctuation properties. An analytical approach must be based
on properties of the matrices denoted by $C_{\mu \nu}(J, \alpha)$ in
this paper. While a theoretical description for shells with several
subshells is probably very difficult, focus on a single $j$--shell
might simplify the problem. (ii) The TBRE predicts correlations
between spectra with different quantum numbers (e.g., different
masses, spins, or isospins) for nuclei within a major shell. The
expermimental verification is difficult due to limitations in the
length and completeness of observed nuclear spectra, but other Fermi
systems might be more accessible. (iii) The correlations between
spectra with different quantum numbers might also affect the
scattering matrix, more precisely: Such correlations might induce
correlations among $S$--matrix elements carrying different total spin
quantum numbers. The present analysis of fluctuating cross sections in
compound nuclei neglects any such correlations. A better understanding
of this problem would be highly desirable.

Nuclear spectroscopy is a mature field with a history of more than 50
years. In spite of this fact it continues to offer great challenges.
We have addressed one of them: The way chaos is induced by the
two--body interaction of the shell model, an interaction which is at
the same time responsible for the many regular features seen in
nuclei.

\begin{acknowledgments}
This work was supported in part by the U.S. Department
of Energy under Contract Nos. DE-FG02-96ER40963 (University of
Tennessee) and DE-AC05-00OR22725 with UT-Battelle, LLC (Oak Ridge
National Laboratory). T.P. thanks the members of the
Max--Planck--Institut f\"ur Kernphysik in Heidelberg for their
hospitality and support.
\end{acknowledgments}


\begin{thebibliography}{99}

\bibitem{Ajz87}J. Ajzenberg-Selove, Nucl. Phys. {\bf A 475}, 1 (1987).

\bibitem{Boh}A. Bohr and B. Mottelson, Nuclear Structure, Vol. 1, W.
A. Benjamin, New York (1969).

\bibitem{Lane}T. Lane and R. Thomas, \rmp {\bf 30}, 257 (1958).

\bibitem{Rain}J. B. Garg, J. Rainwater, J. S. Petersen, and W. W.
Havens Jr., Phys. Rev. {\bf 134}, B985 (1964).

\bibitem{NBohr}N. Bohr, Nature {\bf 137}, 344 (1936).

\bibitem{NBohr1}Nature {\bf 137}, 351 (1936).

\bibitem{Wign}E. P. Wigner, reprinted in: C. E. Porter, Statistical
Theories of Spectra: Fluctuations, Academic Press, New York (1965).

\bibitem{dys}F. J. Dyson, J. Math. Phys. {\bf 3}, 140 (1962); ibid. 164;
ibid. 1199. 

\bibitem{TG}Thomas Guhr, A. M\"uller--Groeling, and H. A. Weidenm\"uller,
Phys. Rep. {\bf 299}, 189 (1998); cond-mat/9707301.

\bibitem{Bohi}O. Bohigas, E. M. Giannoni, and C. Schmit, Phys. Rev. Lett.
{\bf 52}, 1 (1986).

\bibitem{heusler}S. Heusler, S. M\"uller, A. Altland, P. Braun, and F.
Haake, Phys. Rev. Lett. {\bf 98} 044103 (2007); nlin.CD/0610053.

\bibitem{haake}F. Haake, {\it Quantum Signatures of Chaos}, 2nd enlarged
edition, Springer--Verlag, Berlin/Heidelberg (2001). 

\bibitem{Haq}R. V. Haq, A. Pandey, and O. Bohigas, Phys. Rev. Lett. {\bf 48}, 
1086 (1982); and in: Nuclear Data for Science and Technology, K. H.
B\"ockhoff, Editor, Reidel, Dordrecht (1983) 209. 

\bibitem{May}M. Mayer and J. H. D. Jensen, Elementary Theory of Nuclear
Shell Structure, Wiley, New York (1955).

\bibitem{Zel}V. Zelevinsky, B. A. Brown, N. Frazier, and M. Horoi, Phys.
Rep. {\bf 276}, 85 (1996).

\bibitem{Fre}J. B. French and S. S. M. Wong, 
Phys. Lett. {\bf B 33}, 449 (1970).

\bibitem{Boh2}O. Bohigas and J. Flores, 
Phys. Lett. {\bf B 34}, 261 (1971).


\bibitem{Mon}K. K. Mon and J. B. French, 
Ann. Phys. (N.Y.) {\bf 95}, 90 (1975).

\bibitem{Kota}V. B. K. Kota, 
Phys. Rep. {\bf 347}, 223  (2001).

\bibitem{Benet}L. Benet and H. A. Weidenm\"uller, 
J. Phys A: Math. Gen. {\bf 36}, 3569 (2003); cond-mat/0207656.

\bibitem{Pap1}T. Papenbrock and H. A. Weidenm\"uller, Phys. Rev. Lett.
{\bf 93}, 132503 (2004); nucl-th/0404022.

\bibitem{Pap2}T. Papenbrock and H. A. Weidenm\"uller, Nucl. Phys. {\bf 
A 757}, 422 (2005); nucl-th/0403041.

\bibitem{Pap3}T. Papenbrock and H. A. Weidenm\"uller, Phys. Rev. {\bf
C 73}, 014311 (2006); nucl-th/0510018.

\bibitem{adel}A. Y. Abul-Magd, H. L. Harney, M. H. Simbel, and H. A.
Weidenm\"uller, Phys. Lett. {\bf B 579}, 278 (2004); nucl-th/0212057.

\bibitem{Bro}B. A. Brown and B. H. Wildenthal, Ann. Rev. Nucl. Part.
Science {\bf 38}, 29  (1988).

\bibitem{Hon}M. Honma, T. Otsuka, B. A. Brown, and T. Mizusaki,
Phys. Rev. {\bf C 69}, 034335 (2004); nucl-th/0402079. 

\bibitem{Joh}C. W. Johnson, G. F. Bertsch, and D. J. Dean, Phys. Rev.
Lett. {\bf 80}, 2749 (1998); nucl-th/9802066.

\bibitem{Bij}R. Bijker and A. Frank, 
Phys. Rev. Lett. {\bf 84}, 420 (2000); nucl-th/9911067.

\bibitem{Jac}P. Jaquod and A. D. Stone, 
Phys. Rev. Lett. {\bf 84}, 3938 (2000).

\bibitem{Zel1}V. Zelevinsky and A. Volya, 
Phys. Rep. {\bf 391}, 311 (2004); nucl-th/0309071.

\bibitem{Zhao1}Y. M. Zhao, A. Arima, N. Yoshinaga, 
Phys. Rep. {\bf 400}, 1 (2004); nucl-th/0311050.

\bibitem{Bij01}
R. Bijker and  A. Frank,
\prc {\bf 64}, 061303 (2001); nucl-th/0111009.

\bibitem{Zhao2}
Y. M. Zhao, A. Arima, and N. Yoshinaga,
\prc {\bf 66}, 034302 (2002); nucl-th/0112075.

\bibitem{Whe}Y. Alhassid and N. Whelan, 
Phys. Rev. Lett. {\bf 67}, 816 (1991).

\end{thebibliography}
\end{document}